\documentclass[a4paper,11pt]{article}

\usepackage{geometry}

\usepackage[utf8]{inputenc}
\usepackage{amsmath} % for \DeclareMathOperator use below in settings
\usepackage{amssymb} % for math symbols like natural numbers N or real numbers R
\usepackage{subfigure}
\usepackage{comment} % for commenting
\usepackage{dsfont} % support for indicator function
\usepackage{mathrsfs} % e.g. for \LLL
\usepackage{enumerate} % for roman enumaration (i), (ii), ...
\usepackage{outlines} % for easier enumeration \begin{outline} \1 [...] \2 [...] \2 [...] \end{outline}
\usepackage{rotating} % for \includegraphics ???
\usepackage{soul} % for \ul support
\usepackage[round]{natbib} % for using references in the economics style
\usepackage[font = {footnotesize}, labelfont = {bf}, margin = 0.5cm, aboveskip = 5pt, position = bottom]{caption} % not sure what it does, but strange errors when not using it
\usepackage[colorlinks,urlcolor=red,citecolor=blue,linkcolor=red]{hyperref} % for references within paper
\usepackage{xurl}

\usepackage{adjustbox}
\usepackage{setspace}
\usepackage{multirow}

\usepackage{float}

\usepackage{lscape}
\usepackage{algorithm2e} 
\usepackage{algpseudocode} 
\usepackage{tikz}
\usepackage{bm}
\usepackage{booktabs,caption,fixltx2e}
\usepackage{makecell}
\usepackage{threeparttable}
\newcommand{\E}{\mathbb{E}}

\newcommand{\one}{\mathds{1}}

\newcommand{\BOX}{\mbox{{\ensuremath{\Box}}\hspace{-0.5mm}}}

\makeatletter

\renewcommand{\p@enumi}{\thetheorem-}
\@addtoreset{equation}{section}
\makeatother

\numberwithin{figure}{section}%
\numberwithin{table}{section}

\renewcommand\appendix{\par
	\setcounter{section}{0}%
	\setcounter{subsection}{0}%
	\setcounter{figure}{0}%
	\renewcommand\thesection{\Alph{section}}%
	\renewcommand\thefigure{\Alph{section}.\arabic{figure}}}
\newcommand{\LGD}{{\rm LGD}}
\newcommand{\PD}{{\rm PD}}
\newcommand{\ELGD}{{\rm ELGD}}
\newcommand{\VLGD}{{\rm VLGD}}
\newcommand{\MA}{{\rm MA}}
\newcommand{\M}{{\rm M}}

\title{On the Relevance and Appropriateness of Name Concentration Risk Adjustments for Portfolios of Multilateral Development Banks\\[2mm]
}

\author{
	Eva L\"{u}tkebohmert$^1$\footnote{Corresponding author, email: eva.luetkebohmert@finance.uni-freiburg.de}
	\, and Julian Sester$^2$ and Hongyi Shen$^1$
}

\begin{document}
	\onehalfspacing
	\maketitle
	\vspace{-6ex}
	
	\begin{center}
		\small\textit{
			$^{1}$Department of Quantitative Finance,\\
			Institute for Economic Research, University of Freiburg,\\
			Rempartstr. 16,	79098 Freiburg, Germany.\\[2mm]
		$^{2}$National University of Singapore, Department of Mathematics,\\ 21 Lower Kent Ridge Road, 119077 Singapore}            
	\end{center}

	\begin{abstract}
    Sovereign loan portfolios of Multilateral Development Banks (MDBs) typically consist of only a small number of borrowers and hence are heavily exposed to single name concentration risk. Based on realistic MDB portfolios constructed from publicly available data, this paper quantifies the magnitude of the exposure to name concentration risk using exact Monte Carlo simulations. In comparing the exact adjustment for name concentration risk to its analytic approximation as currently applied by the major rating agency Standard \& Poor's, we further investigate whether current capital adequacy frameworks for MDBs are overly conservative. Finally, we discuss the choice of appropriate model parameters and their impact on measures of name concentration risk. 
	\end{abstract}

\section{Introduction}

Multilateral Development Banks (MDBs) raise a large volume of funds on international capital markets to finance projects that mitigate climate change and promote social development and economic growth in developing countries. The important role of MDBs in development finance, in particular in light of the 2030 Agenda for Sustainable Development is pointed out in \cite{UN2015}, \cite{Griffith2016} and \cite{Gurara2020}.
In this sense, MDB debt is increasingly relevant as a specialized sub-component of fixed income and can be considered as a sub-set of the broader social/green bond space. 
As non-profit supranational development institutions MDBs are not regulated and hence there are no generally agreed standards to orient stakeholders. Given their financial model it is therefore essential for MDBs to maintain strong credit ratings, which are, however, highly dependent on how credit rating agencies (CRAs) assess MDBs' capital adequacy.
An important risk component in MDBs loan portfolios is their exposure to single name concentration risk which arises from the fact that MDBs' development-related lending is mainly sovereign lending, so that their loan portfolios typically consist of a small number of borrowers.
The leading methodology that is currently in use (e.g.\ in \cite{S&P2018}) to account for name concentration risk in MDBs' capital adequacy framework relies on the Granularity Adjustment (GA) developed in \cite{GordyLuetkebohmert2013}. The latter was originally designed for commercial banks which typically hold much larger portfolios consisting of at least several hundred borrowers.

In this paper, we investigate the degree to which MDBs are exposed to single name concentration risk and we analyse whether the GA methodology currently applied by Standard \& Poor's (S\&P) is overly conservative when applied to MDB portfolios. 
The exact measure of name concentration risk in a portfolio can be quantified as the difference between the Value-at-Risk (VaR) of the true portfolio loss variable and the VaR of an analogous asymptotic portfolio where all idiosyncratic risk is diversified away (i.e.\ a perfectly fine grained portfolio). When applying a single-factor credit portfolio model as e.g.\ the Vasicek model, the latter can be expressed as the conditional expectation of the portfolio loss given the quantile of the single systematic risk factor.
The GA methodology is an asymptotic approximation to the true measure of name concentration risk and has been shown to be very accurate for medium and larger commercial bank portfolios. However, as an analytic approximation its accuracy decreases with the number of borrowers which may lead to substantial approximation errors when portfolios of less than one hundred obligors are considered. Furthermore, the GA in \cite{GordyLuetkebohmert2013}, which is used under the S\&P capital adequacy framework for MDBs, builds on the CreditRisk$^+$ model which applies a Poisson approximation to the conditional default probabilities in order to derive an analytic representation of the probability generating function (PGF) of the portfolio loss distribution. While this is well motivated when considering commercial bank portfolios which typically have borrowers of relatively high creditworthiness, the approximation becomes less accurate when the default probabilities of borrowers increase. This might be a serious concern for MDBs whose borrowers are usually much lower rated. Moreover, relying on an actuarial definition of loss, the methodology ignores effects from rating transitions and does not sufficiently account for loan maturities. 
In addition, the approach in \cite{S&P2018} can be criticized from a practical point of view since it applies the GA formula with exactly the same parameter values as in \cite{GordyLuetkebohmert2013}. These parameters were, however, calibrated to data on commercial bank portfolios and hence might not be appropriate for portfolios of MDBs. In defense of the S\&P approach, it has to be acknowledged, however, that S\&P has put a boundary on the use of the GA formula by imposing a floor of B- on the sovereign ratings in order to compensate for a potential overestimation of name concentration risk when applying the GA methodology to MDBs.
%Regulatory arbitrage refers to a structuring of banking  activities such that regulatory requirements are reduced without a corresponding reduction in the underlying risk.

This paper provides a comprehensive empirical study of name concentration risk in MDB portfolios based on realistic portfolios constructed from financial statements of eleven MDBs in 2022. More specifically, we first evaluate the importance of name concentration as risk factor in MDB portfolios based on an exact measurement using Monte Carlo (MC) simulations. We then compare the analytical approximation GA, as currently implemented under the approach in \cite{S&P2017,S&P2018}, to the exact GA calculations obtained through MC simulations. Since the GA formula in \cite{GordyLuetkebohmert2013} is calibrated to the Internal Ratings Based (IRB) approach of the Basel regulatory framework, we compare it to MC simulations of the exact GA in an actuarial version of the one-factor CreditMetrics model that underpins the IRB model. Additionally, we study the effect of rating transitions and loan maturities on the exact and approximate GA measures by considering also a Mark-to-Market (MtM) CreditMetrics model. To this end, we build on the analytic GA formula for the MtM setting as developed in \cite{GordyMarrone2012} as well as on corresponding exact GA calculations through MC simulations. 
Further, we analyse the impact of different parameter specifications that may be substantially different for MDBs compared to commercial banks. Finally, we acknowledge that the capital adequacy framework of \cite{S&P2018} allows for adjusted rating transition probabilities and loss given default (LGD) rates due to MDBs' preferred creditor treatment (PCT). We analyse how these PCT-adjusted inputs affect the exact and approximate GAs. 

Our main findings document that MDBs are indeed severely exposed to name concentration risk with exact GAs of up to 38\% of total portfolio exposure, accounting for up to 82\% of total unexpected loss. Further, our results indicate that the current methodology applied by the rating agency is indeed overly conservative. Approximate GAs overestimate the exact GA in MDB loan portfolios by up to 28 percentage points for constant loss given default (LGD) rates and up to 37 percentage points when LGDs are random. In relative terms, the approximate GA can lead to an increase of 266\% of the exact GA for some of the MDB portfolios. In addition, we show that PCT has a very strong impact on the size of the GA. Adjusting LGD rates and transition probabilities for PCT can lead to significant reductions in GAs when measured in percentage of total portfolio exposure whereas relative GAs as ratio of total unexpected loss stay at high levels. 
%Furthermore, our results show that the choice of the model parameters as currently applied under the S\&P approach is inappropriate. When calibrating these parameters to MDB portfolio data, the GA decreases by...

While our analysis mainly focuses on the GA method as implemented in \cite{S&P2018}, it has to be noted that the three major credit rating agencies each take different approaches to incorporating concentration risk into their assessment of MDB capital adequacy. Unlike S\&P, Moody's and Fitch follow a more qualitative approach. 
Moody's assessment of the credit risk for MDBs relies on a score card approach using both qualitative and quantitative indicators (e.g.\ for leverage, development asset credit quality, liquidity, funding structure, quality of risk management, shareholders' ability to support the MDB, etc.), which may be evaluated based on historical or forward-looking data.
The credit quality of the development assets is assessed qualitatively and also takes into account portfolio concentrations by constraining the score for the factor. In terms of single name concentration risk, it mainly takes into account the percentage of exposure to the largest borrower as well as to the top 10 borrowers in the portfolio, see \cite{Moodys2020}. The rating methodology of Fitch is comparable. Starting from an intrinsic rating that depends on liquidity and solvency factors involving both quantitative and qualitative indicators, adjustments are made for MDB's governance and business environment. The solvency factor also depends on the concentration risk in MDB loan portfolios which is measured by the ratio of the five largest exposures to the total banking portfolio, see \cite{Fitch2020}. While these approaches might result in a lower penalty for name concentration risk than the S\&P approach, they also do not provide an accurate quantitative measure of name concentrations in loan portfolios. 

Our paper relates to the large literature on name concentration risk. The GA methodology as suggested in \cite{Gordy2003} and \cite{Wilde2001} is an adjustment to the Asymptotic Single Risk Factor (ASRF) model underpinning the Basel regulatory framework and was introduced in an earlier version of Basel II known as the Second Consultative Paper (see \cite{Basel2001CP2}). \cite{MartinWilde2002} provide a mathematical rigorous derivation of the GA formula building on theoretical work by \cite{Gourieroux}.  Other earlier contributions are due to \cite{PykhtinDev2002} and \cite{Wilde2001IRB}. \cite{EmmerTasche2004} suggest a GA based on a single-factor default-model CreditMetrics model while \cite{GordyMarrone2012} develop a GA for a MtM credit portfolio model. The GA in \cite{GordyLuetkebohmert2013}, in contrast, is based on the actuarial CreditRisk$^+$ model but is calibrated to the inputs of the IRB model which makes it very attractive from a practical perspective. \cite{EbertLuetkebohmert2011} extend the methodology to loan portfolios with hedged positions. The accuracy of the GA formula for commercial bank portfolios has been verified also in empirical work of \cite{TarashevZhu2008}, \cite{Heitfieldetal2006}) and \cite{GuertlerHeitheckerHibbeln}. While all the above studies focus on commercial bank portfolios, the application of the GA methodology to MDB portfolios as currently implemented under the S\&P approach has been questioned already in works by \cite{Perraudin2016} and \cite{Humphrey2015,Humphrey2018}. Our paper contributes to this discussion in providing a comprehensive and rigorous analysis of name concentration risk in MDB portfolios by comparing the exact measure of name concentration risk derived from MC simulations with the approximate GA calculations for several realistic MDB portfolios. In this way, it also relates to work by \cite{Humphrey2017} who discusses how credit agencies' evaluation of the financial strength of MDBs affects their lending headroom and thereby influences MDBs' operations to achieve their development goals. It is also specifically recommended in \cite{CAFReview2022} that MDB capital adequacy frameworks should appropriately account for PCT and concentration risk in MDB portfolios and that rating agencies' methodologies should be further evaluated in this respect (compare Recommendations 1B on p. 28 and 4 on p. 41). Our study accomplishes exactly this need.

The paper is structured as follows. Section \ref{sec methodology} reviews the methodology to measure name concentration risk. Here we discuss the exact GA based on MC simulation in both the one-factor actuarial CreditMetrics model and the MtM CreditMetrics model and we outline the approximation GA developed in \cite{GordyLuetkebohmert2013}. Section \ref{sec data} describes the construction of the realistic MDB portfolios based on publicly available data. Section \ref{sec results} discusses our numerical results while Section \ref{sec conclusion} summarizes and concludes.

\section{Methodology}\label{sec methodology}

In this section, we first discuss the exact measure of name concentration risk in the Basel regulatory framework and give a very brief review of the GA methodology as implemented in \cite{S&P2017,S&P2018}. We then discuss the theoretical and practical challenges associated with the analytic GA when applied to MDB portfolios. Finally, we explain the exact GA calculation in the MtM CreditMetrics model.

\subsection{Exact GA in the Basel Regulatory Framework}\label{sec GA IRB model}

Consider a portfolio of $N$ borrowers and denote the portfolio loss rate by
\begin{equation}\label{eq_defn_L_actuarial}
L=\sum_{n=1}^N a_n \LGD_n D_n,
\end{equation} 
where $a_n=A_n/\sum_{i=1}^N A_i$ is the exposure $A_n$ to borrower $n$ as a share of total portfolio exposure and $\LGD_n$ denotes the loss given default associated with borrower $n$ which can be deterministic or stochastic. The default indicators $D_n$ are assumed to be conditionally independent across borrowers given a systematic risk factor $X$. 

The exact adjustment for single name concentration risk in a loan portfolio is the difference between the true portfolio value-at-risk (VaR) $\alpha_q(L)$ and the VaR of an infinitely fine-grained portfolio 
\begin{equation}\label{equ exact GA general}
GA_q^{\rm exact}(L)\equiv \alpha_q(L)- \mathbb{E}[L|\alpha_q(X)].
\end{equation}
Under the actuarial definition of portfolio loss (\ref{eq_defn_L_actuarial}) the exact GA can be expressed as
\begin{equation}\label{equ GA exact actuarial}
    GA^{\rm exact, act}_q(L)= \alpha_q\left(\sum_{n=1}^N a_n \LGD_n  D_n\right)- \sum_{n=1}^N a_n \ELGD_n \pi_n(\alpha_q(X)).
\end{equation}
where $\ELGD_n$ denotes the expected LGD and $\pi_n(x)$ is the conditional default probability given a realization $X=x$ of the systematic risk factor. We assume the borrower specific $\LGD_n$ to be beta distributed with mean $\ELGD_n$ and variance $\VLGD_n^2=\nu  \ELGD_n(1-\ELGD_n)$ for some $\nu\in[0,1]$. To make expression (\ref{equ GA exact actuarial}) more explicit, the distribution of the systematic risk factor and the default indicator need to be specified. 

In the ASRF model underpinning the Basel IRB approach, the default indicator $D_n=\one_{\{Y_n\leq C_n\}}$ equals one when the latent asset return 
\begin{equation}\label{equ latent asset return}
Y_n=\sqrt{\rho_n} X+\sqrt{1-\rho_n} \epsilon_n
\end{equation}
of borrower $n$ drops below a certain threshold $C_n=\Phi^{-1}(\PD_n)$ and is zero otherwise. Here $\Phi$ denotes the standard normal cumulative distribution function and $\PD_n$ is borrower $n$'s unconditional default probability. The single systematic risk factor $X$ and the idiosyncratic risk factors $\epsilon_n$ are independent and standard normally distributed.
The parameter $\rho_n$ is the asset correlation and is specified in the IRB approach as a function of PD, i.e.\
\begin{equation}\label{equ IRB asset correlation}
    \rho_n=0.12\cdot \frac{1-e^{-50\cdot \PD_n}}{1-e^{-50}}+0.24\cdot \left(1-\frac{1-e^{-50\cdot \PD_n}}{1-e^{-50}}\right).
\end{equation}
The conditional default probability in (\ref{equ GA exact actuarial}) can be expressed as
\begin{equation}\label{equ cond PD Vasicek model}
  \pi_n(\alpha_q(X))=  \Phi\left(\frac{\Phi^{-1}(\PD_n)+\sqrt{\rho_n}\Phi^{-1}(q)}{\sqrt{1-\rho_n}}\right).
\end{equation}
Under some technical assumption (compare \cite{Gordy2003}), in this model the second term in (\ref{equ GA exact actuarial}) representing the conditional expected loss $\mathbb{E}[L|\alpha_q(X)]$ given the $q^{\rm th}$ quantile of $X$ converges to the first term $\alpha_q(L)$ when the number $N$ of borrowers increases. Thus, the exact GA vanishes for infinitely fine grained portfolios. For smaller portfolios, however, the GA is non-negligible.

The above specification can be seen as a one-factor actuarial version of the CreditMetrics model which resembles the single-factor Mark-to-Market (MtM) Vasicek model from which the IRB formula is actually derived. The exact GA (\ref{equ GA exact actuarial}) can be determined by MC simulation, i.e.\ by simulating the latent asset returns $Y_n$ according to (\ref{equ latent asset return}) for standard normal factors $X$ and~$\epsilon_n$. 
%Since a MtM approach poses considerable additional challenges which might affect our comparison study, we rather use the above actuarial version as benchmark approach in our analysis. 

\subsection{GA Methodology Applied by S\&P}\label{sec analytic GA}

The leading methodology for the quantification of name concentration risk in MDB portfolios currently applied in the capital adequacy frameworks of the rating agency S\&P is based on the GA formula derived in \cite{GordyLuetkebohmert2013}. The latter is based on a CreditRisk$^+$ setting, where the systematic risk factor $X$ in the general formula (\ref{equ GA exact actuarial}) is Gamma distributed with mean 1 and variance $1/\xi$ for some $\xi>0$ and the conditional default probabilities equal
\begin{equation}\label{equ cond PD actuarial}
\pi_n(x)=\PD_n (1+\omega_n(x-1)),
\end{equation}
with $\PD_n$ denoting the unconditional default probability of borrower $n$ and factor weights $\omega_n$ specifying the sensitivity of borrower $n$ to the systematic risk factor $X$.

The GA derived in \cite{GordyLuetkebohmert2013} is an analytic approximation to the exact GA in (\ref{equ GA exact actuarial}) and can be expressed as
\begin{equation}\label{equ GA 1st order actuarial}
\begin{array}{ccl}
GA_q^{\operatorname{approx, act}}&=&\displaystyle  \frac{1}{2\mathcal{K}^*} \sum_{n=1}^N a_n^2 \Big[\delta\left(\mathcal{C}_n (\mathcal{K}_n+\mathcal{R}_n)+(\mathcal{K}_n+\mathcal{R}_n)^2 \frac{\VLGD^2_n}{\ELGD_n^2}\right)\\
&&\displaystyle\hspace{1cm}  - \mathcal{K}_n \left(\mathcal{C}_n +2(\mathcal{K}_n+\mathcal{R}_n) \frac{\VLGD^2_n}{\ELGD_n^2}\right)\Big]
\end{array}
\end{equation}
where $\mathcal{C}_n=\frac{\VLGD_n^2+\ELGD_n^2}{\ELGD_n}$ and 
\begin{equation}\label{equ VLGD}
\VLGD_n^2=\nu\cdot \ELGD_n(1-\ELGD_n)
\end{equation}
for some parameter $\nu\in[0,1]$.
The inputs to the GA are expressed in terms of the EL reserve requirement $\mathcal{R}_n=\ELGD_n \PD_n$ and the UL capital requirement $\mathcal{K}_n$ as calculated under the Basel IRB formula, i.e.\ 
$$
\mathcal{K}_n=\left( \ELGD_n\cdot \Phi\left(\frac{\Phi^{-1}(\PD_n)+\sqrt{\rho_n}\Phi^{-1}(q)}{\sqrt{1-\rho_n}}\right)-\PD_n\cdot \ELGD_n\right) \cdot \MA_n,
$$
where $\rho_n$ is the asset correlation as in (\ref{equ IRB asset correlation}) and $\MA_n$ is a maturity adjustment factor for the loan of maturity $\M_n$ to borrower $n$ given by
\begin{equation}\label{equ MA factor}
    \MA_n=\frac{1 + (\M_n - 2.5) \cdot b (\PD_n)}{1 - 1.5 \cdot b(\PD_n)}\;\mbox{with}\; b(\PD_n) = (0.11852-0.05478\cdot \log(\PD_n))^2.
\end{equation}
The maturity adjustment equals one for $\M_n=1$ and is larger than one when the maturity $\M_n>1.$
Moreover, $\mathcal{K}^*=\sum_{n=1}^N a_n \mathcal{K}_n$ and 
$$
\delta\equiv-(\alpha_q(X)-1) \frac{h'(\alpha_q(X))}{h(\alpha_q(X))}=(\alpha_q(X)-1)\left(\xi+\frac{1-\xi}{\alpha_q(X)}\right).
$$
By ignoring higher order terms in $\PD$s the GA can be simplified further as
\begin{equation}\label{equ GA 1st order actuarial simplified}
\begin{array}{ccl}
GA_q^{\operatorname{simplified}}&=&\displaystyle  \frac{1}{2\mathcal{K}^*} \sum_{n=1}^N a_n^2 \Big[\mathcal{C}_n \left(\delta(\mathcal{K}_n+\mathcal{R}_n)- \mathcal{K}_n \right)\Big].
\end{array}
\end{equation}
The capital adequacy framework for MDBs in \cite{S&P2018} applies this simplified GA for measuring name concentration risk where in line with \cite{GordyLuetkebohmert2013} the volatility parameter is set to $\nu=0.25$, the precision parameter is $\xi=0.25$, and the asset correlation is specified as in the IRB approach (\ref{equ IRB asset correlation}), which maps the borrower PD to an asset correlation between 12\% and 24\%.

%While the above analytical approximation is very accurate for larger commercial bank portfolios,
%it may produce less reliable results for small portfolios as those typically held by MDBs. 

\subsection{Drawbacks of Existing Methodology}\label{sec drawbacks}

Before turning to the empirical evaluation of the exact measure of name concentration risk in MDB portfolios and the appropriateness of the approximate GA to account for this risk source, we first discuss some theoretical challenges of the approximate GA when applied to MDB loan portfolios. 

The GA is an analytical approximation to the exact measure of name concentration risk. As documented in \cite{GordyLuetkebohmert2013} (see also \cite{TarashevZhu2008}) the approach is very accurate for commercial bank portfolios with several hundred borrowers. However, the approximation may perform considerably worse if the number $N$ of borrowers in the portfolio becomes very small. In fact, when considering the extreme case of only a single loan, we obtain that if $\PD$ is greater than one minus the VaR threshold~$q$ and $\LGD$ is fixed, then the VaR equals $\LGD$, which for plausible values of~$\rho$ is substantially larger than the systematic risk component $\E[L|\alpha_q(L)]$.

Moreover, the GA in (\ref{equ GA exact actuarial}) builds on the CreditRisk$^+$ framework which approximates the conditional distribution of the default indicator variables given the risk factor $X$ by a Poisson distribution in order to maintain the analytical tractability of the portfolio loss distribution. The approximation error is proportional to the squared default probability and hence very small when the borrowers in the loan portfolio have very low default risk. However, the average ratings of sovereign borrowers in MDB portfolios typically range between BB+ and CCC+, so that there are also several loans which have a rather high associated default risk making the Poisson approximation less reliable.

Similarly to the above issue, the simplified GA ignores terms which are of quadratic (or higher) order in PDs. This again may lead to a non-negligible approximation error when applied to MDB portfolios. While this problem could be circumvented by applying the full analytic approximation GA (\ref{equ GA 1st order actuarial}) including higher order terms in PDs, the other two problems still remain.

In addition, from a practical perspective the choice of the parameters is disputable since these where calibrated to commercial bank data. A recent study by \cite{RiskControl2023}, for example, indicates a substantially higher asset correlation for borrowers in MDB portfolios. 

Finally, the GA method is based on an actuarial definition of loss and hence ignores effects from rating transitions. Further, it does not sufficiently account for loan maturity since the maturity adjustment in (\ref{equ MA factor}) is a rather rough approximation obtained by a smoothed regression approach. \cite{GordyMarrone2012} have extended the GA methodology to a MtM definition of loss and derive an explicit solution for the MtM CreditMetrics ratings-based model. While they show that the MtM GA does depend on rating transitions and loan maturities, as an asymptotic approximation their approach, however, also suffers from the problems stated above. Nevertheless, to address the effect of rating transitions and loan maturities on the GA, we include their approximate MtM GA formula as well as the exact GA in the MtM CreditMetrics model, as described in the following subsection, in our empirical study.

\subsection{Exact GA in Mark-to-Market CreditMetrics Model}

In this section, we introduce the exact GA in the MtM CreditMetrics model which allows an in-depth analysis of the effect of varying loan maturities on the GA. Further, the model can be used to calibrate the free model parameter $\xi$ in the CreditRisk$^+$ GA.

In the MtM setting, the loss $L_n$ on position $n$ is defined as the difference between the expected return $\mathbb{E}[R_n]$ and the realized return $R_n$, discounted to the current date at the (continuously compounded) riskfree interest rate~$r$, where return $R_n$ is defined as the ratio of market value at time $T$ to the current time $0$ market value. Hence, the total portfolio return is $R=\sum_{n=1}^N a_n R_n$ and the total portfolio loss rate equals
\begin{equation}\label{equ portfolio loss MtM}
L=\left(\mathbb{E}[R]-R\right)\cdot e^{-rT}.
\end{equation}
We consider a ratings-based approach analogous to the CreditMetrics model. Hence, at the final time horizon $T$ each obligor $n$ is assigned a rating $S_n\in\mathcal{S}=\{0,1,\ldots, S\},$ where state $0$ is the absorbing default state and state $S$ denotes the highest possible rating (e.g.\ AAA rating in S\&P's rating classification scheme).
Denote by $p_{gs}$ the transition probability from current rating $g$ to rating $s$ at time $T$. 
We associate to each borrower $n$ a latent asset return 
$$
Y_n=\sqrt{\rho_n} X+\sqrt{1-\rho_n} \epsilon_n,
$$
with a standard normally distributed systematic risk factor $X$ and independent standard normally distributed idiosyncratic risks $\epsilon_n$. Here $\rho_n$ denotes the asset correlation of borrower $n$. The conditional probability that obligor~$n$ is in state $S_n=s$ at time $T$ given $X=x$ then equals
\begin{equation}\label{eq_pi_def}
\pi_{ns}(x)=\Phi\left(\frac{C_{g(n),s}-x\sqrt{\rho_n}}{\sqrt{1-\rho_n}}\right)-\Phi \left(\frac{C_{g(n),s-1}-x\sqrt{\rho_n}}{\sqrt{1-\rho_n}}\right),
\end{equation}
where $C_{g(n),s}$ and $C_{g(n),s-1}$ denote the threshold values so that borrower $n$ with current rating $g(n)$ is in rating $s$ at time $T$ when $C_{g(n),s-1}<Y_n\leq C_{g(n),s}$, which can be derived from the transition probabilities as
$$
C_{g,s}=\Phi^{-1}\left(\sum_{i=0}^s p_{gi}\right),\quad\mbox{for}\quad s=0,\ldots,S-1.
$$

Following \cite{GordyMarrone2012} position $n$ is modelled as a bond with face value 1, maturity in $\tau>T$ years, coupon payments of $c_n \delta$ at times $t_1,\ldots,t_{m}=\tau$ with accrual period $\delta=t_i-t_{i-1}$ (expressed as fraction of year), and current market value $P_{n0}$.
The return on position $n$ can then be expressed as
\begin{equation}\label{equ return R_n}
\begin{aligned}
    R_n&= \sum_{s=1}^S \frac{P_{nT}(s)}{P_{n0}}\cdot\one_{\left\{\Phi^{-1}\left(\sum_{i=0}^{s-1} p_{g(n),i}\right)< ~Y_n~\leq \Phi^{-1}\left(\sum_{i=0}^{s} p_{g(n),i}\right)\right\}},
\end{aligned}
\end{equation}
where $P_{nT}(s)$ denotes the value of the bond $n$ at time $T$ conditional on state $S_n=s$ (see \cite{GordyMarrone2012} for an explicit expression of this quantity).

Given a realization $X=x$ of the systematic risk factor the conditional expectation of the individual and total portfolio return are then equal to
\begin{equation}\label{equ expected return given x}
\begin{array}{ccl}
\mu_n(x)\equiv \mathbb{E}[R_n|X=x]&=&\sum_{s=0}^S \frac{P_{nT}(s)}{P_{n0}}\cdot \pi_{ns}(x)\quad \mbox{and}\quad\\[2mm]
    \mu(x)\equiv\mathbb{E}[R|X=x]&=&\sum_{n=1}^N a_n \mu_n(x),
\end{array}
\end{equation}
where we assume that the market credit spreads at maturity $T$ are solely functions of the rating (independent of the realization of the risk factor $X$).

In the MtM setting $-X$ is the systematic risk factor. Further, we note that the expected return $\mathbb{E}[R]$ in the expression of the portfolio loss rate (\ref{equ portfolio loss MtM}) cancels out when computing the exact GA (\ref{equ exact GA general})
and we obtain the following expression for the exact GA in the MtM setting

\begin{align}
&GA^{\rm exact, MtM}_q(L)= \alpha_q(L)- \E[L~|~\alpha_{q}(-X)]  = \alpha_q(L)- \E[L~|~\alpha_{1-q}(X)]\label{equ GA exact MtM} \\
    =& e^{-rT}\bigg[ \sum_{n=1}^N a_n \cdot \mu_n(\alpha_{1-q}(X)) +\alpha_q\left(-\sum_{n=1}^N a_n \sum_{s=0}^S \frac{P_{nT}(s)}{P_{n0}}\cdot\one_{\left\{C_{g(n),s-1}< ~Y_n~\leq C_{g(n),s}\right\}}\right) \bigg].\nonumber
\end{align}

In Section \ref{sec results}, we evaluate the accuracy of the approximate GA in (\ref{equ GA 1st order actuarial}) of \cite{GordyLuetkebohmert2013}, its simplified version (\ref{equ GA 1st order actuarial simplified}) as implemented in \cite{S&P2018}, and the MtM GA of \cite{GordyMarrone2012} by comparing them against the exact GAs (\ref{equ GA exact actuarial}) and (\ref{equ GA exact MtM}) as derived in both the actuarial and the MtM CreditMetrics models for a set of realistic MDB portfolios constructed from publicly available data.
Since the approximate GA  in (\ref{equ GA 1st order actuarial}) has been calibrated to the IRB inputs, we benchmark it against the exact GA in the model underpinning the IRB approach rather than against the exact GA in the CreditRisk$^+$ model where also the factor weights $\omega_n$ need to be specified which might introduce some additional source of model risk. 

\section{Construction of Realistic MDB Portfolios}\label{sec data}

We extract portfolio data from MDBs' financial statements as of 2022. These contain information on sovereign loan exposures as well as some aggregate data on loan maturities. We combine this with data on sovereign ratings provided by different rating agencies and build on rating transitions as published in \cite{S&P2022} to infer sovereign default probabilities. In the following we provide detailed information on the construction of the realistic loan portfolios for the following MDBs:

\begin{itemize}
    \item African Development Bank (AfDB); see \cite{AfDB2022}
    \item Asian Development Bank (ADB); see \cite{ADB2022}
    %\item Asian Infrastructure Investment Bank (AIIB)
    %\item Bank of the South
    \item Development Bank of Latin America and the Caribbean (CAF); see \cite{CAF2022}
    \item Caribbean Development Bank (CDB); see \cite{CDB2022}
    \item Central American Bank for Economic Integration (CABEI); see \cite{CABEI2022}
    %\item Development Bank of the Central African States (BDEAC)
    \item East African Development Bank (EADB); see \cite{EADB2022}
    %\item Eurasian Development Bank (EDB)
    \item European Bank for Reconstruction and Development (EBRD); see \cite{EBRD2022}
    %\item European Investment Bank (EIB)
    \item Inter-American Development Bank (IDB); see \cite{IDB2022} %sometimes IADB
    %\item International Development Finance Club (IDFC)
    \item International Bank for Reconstruction and Development (IBRD); see \cite{IBRD2022}
    %\item International Development Association (IDA)
    %\item Islamic Development Bank (IsDB)
    %\item New Development Bank (NDB)
    %\item Nordic Investment Bank (NIB)
    \item Trade and Development Bank (TDB); see \cite{TDB2022}
    \item West African Development Bank (BOAD); see \cite{BOAD2022}
\end{itemize}

\begin{figure}[htb]
\begin{center}
       \includegraphics[width=0.8\textwidth]{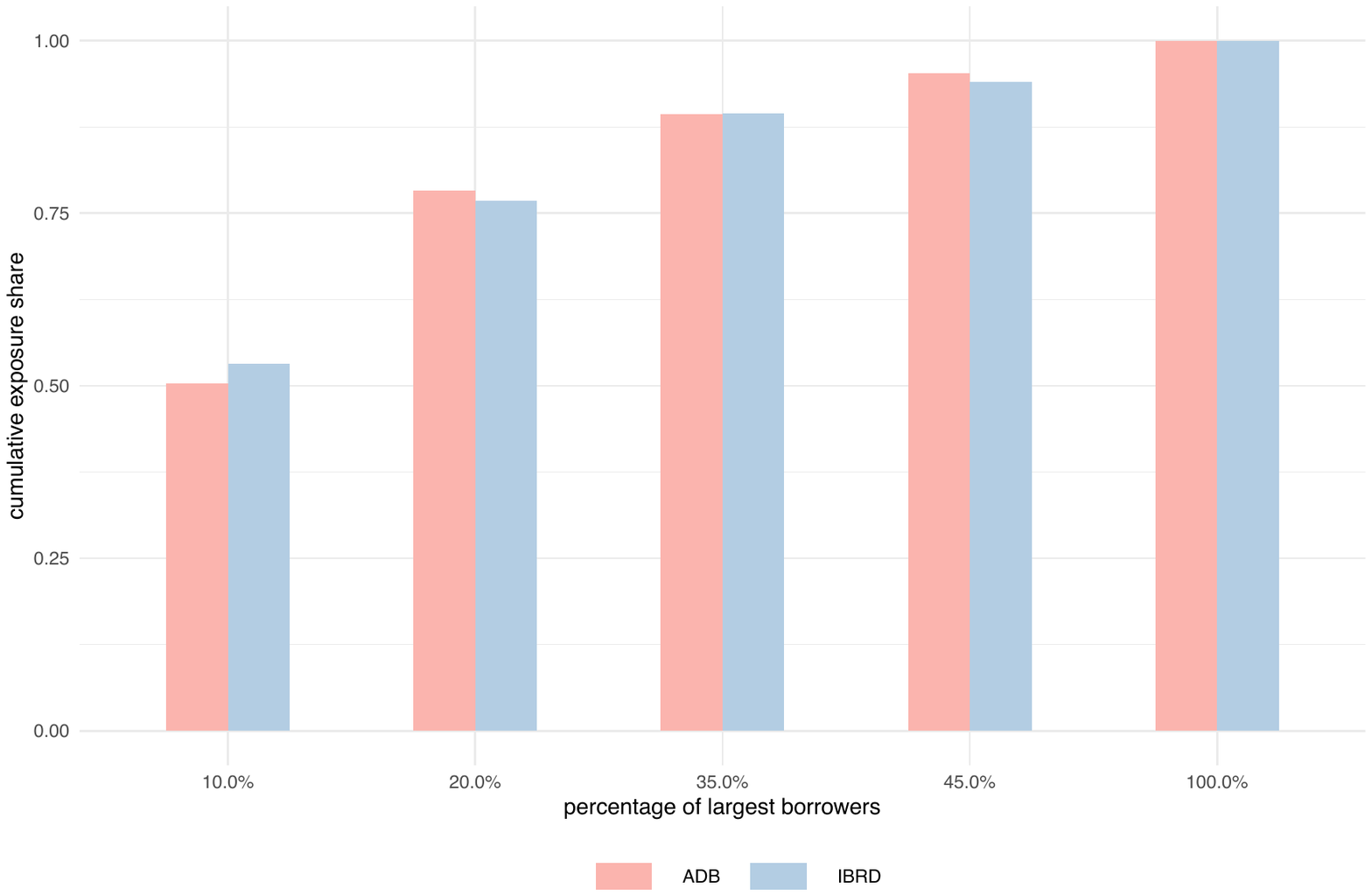}
\caption{Exposure distribution in MDB portfolios. The figure shows the exposure distribution of the ADB and IBRD sovereign loan portfolio as of 2022.} \label{fig exposure distribution ADB IBRD}
\end{center}
\end{figure}

To measure name concentration risk in loan portfolios all loans first need to be aggregated on borrower level. We denote by $A_n$ the total notional amount outstanding for borrower $n$ as quoted in the banks' financial statement (including also the face value of undisbursed loans) and we denote by $a_n$ its share as a fraction of total portfolio exposure. Table \ref{tab portfolios summary} reports the number of borrowers and total exposure for each MDB in our data set for end of 2022. Exposures are expressed in millions of USD. When loans are reported in a different currency -- this is the case for AFDB, EBRD, and BOAD -- we convert these to USD using the exchange rate at the time of the financial statement. Note also that some MDBs have loans to ``regionals'' (e.g.\ ADB). We exclude these since our GA calculations also require borrower specific PDs which are not available for regionals. Moreover, we exclude all non-sovereign loans in the portfolios. Figure \ref{fig exposure distribution ADB IBRD} shows the exposure distribution for the ADB and IBRD portfolios as of 2022. It indicates that approximately 50\% of the portfolio exposure is concentrated on only the largest 10\% of borrowers and more than 75\% of the portfolio exposure is concentrated on the largest 20\% of borrowers. This documents that the MDB portfolios are not only highly concentrated because they are small in terms of the number of borrowers but also because a large fraction of the total exposure is concentrated on a small portion of borrowers. Exposure distributions for the other MDBs in our sample are depicted in Figure \ref{fig exposure distributions other MDBs} in the \ref{app supplementary material}. \\
%We could also calculate the total EAD by applying a Credit Conversion Factor (CCF) of 50\% as in \cite{RiskControl2023} to the face value of undisbursed loans, i.e.\ Total EAD= Outstanding Loans +50\% x Undisbursed Loans. 

\begin{figure}[htb]
\begin{center}
       \includegraphics[width=\textwidth]{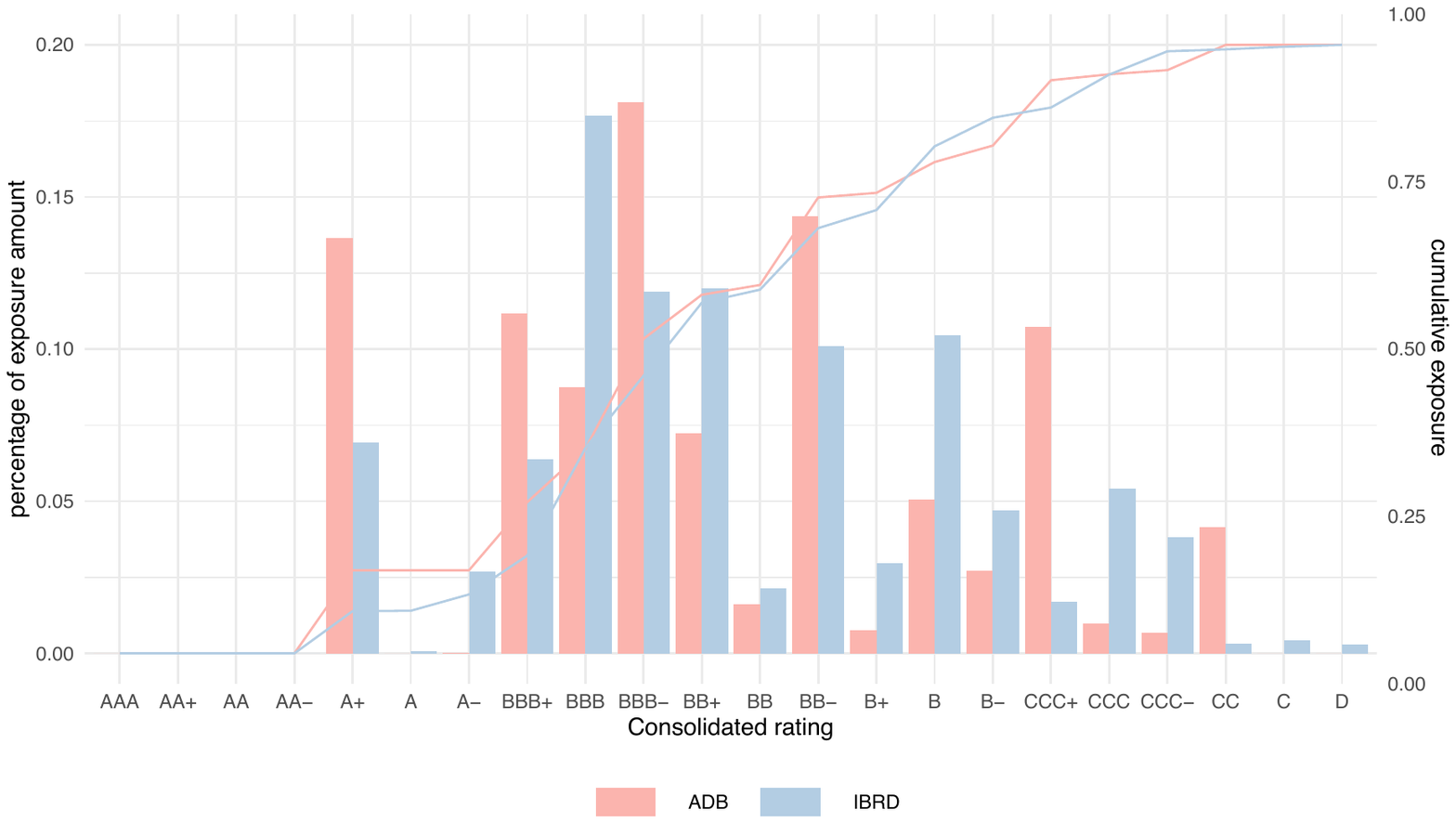} 
\caption{Rating distribution of MDB portfolios. The figure shows the rating distribution of the ADB and IBRD sovereign loan portfolio as of 2022.} \label{fig PD ADB and IBRD}
\end{center}
\end{figure}

We determine credit ratings for individual borrowing countries based on S\&P, Moody's and Fitch ratings. For countries that are not rated by these agencies (approx. 27\% of countries) we infer ratings provided by the Organisation for Economic Co-operation and Development (OECD) and convert these to the rating scale of the major rating agencies by regressing the S\&P ratings on OECD ratings (approx. 17\% of all countries; compare also \cite{RiskControl2023} for this approach). For countries that are also not included in the OECD ratings, we either use the rating quoted on wikirating (3\% of countries) or (if not available) we assign a minimum rating of B- (7\% of countries). This is in line with the S\&P approach where a minimum rating around B- is assigned for unrated countries (as pointed out by MDB officials, compare also footnote 23 in \cite{Humphrey2015}). Figure \ref{fig PD ADB and IBRD} illustrates the rating distribution of the ADB and IBRD loan portfolio as of 2022. Rating distributions for other MDBs are comparable (compare also Figure \ref{fig rating distributions other MDBs} in the Appendix).

%\item Take latest financial statements and corresponding S\&P rating reports so that the dates of exposures and sovereign ratings agree. 

To convert ratings to default probabilities we use the average one-year foreign currency rating transition matrix estimated for the time period 1975--2021 as published in Table 35 by \cite{S&P2022}. We use the foreign currency transition matrix since most loans to sovereign borrowers of various countries are denominated in USD or EUR (see e.g.\ \cite{IBRD2021}). In the financial statements of MDBs some borrowing countries are listed as ``in default''. However, \cite{RiskControl2023} points out that these sovereigns are typically not in default to MDBs due to their preferred creditor status. Hence, we follow their approach and merge all ratings CCC+ and worse into a consolidated rating `Cs', i.e.\ we add up all rating transition probabilities for ratings CCC+ up to CC in the transition matrix of \cite{S&P2022}. 
%This is more conservative than the approach in \cite{S&P2017,S&P2018} where a minimal rating of around B- is imposed. 
Further, since \cite{S&P2022} also consider transitions to a non-rated (`NR') category which we omit, we normalize their transition matrix by dividing all entries by 1 minus the NR transition probability. Table \ref{tab transition matrix} reports the normalised transition matrix which is also used in \cite{RiskControl2023} (see Table A3.5).
To study the impact of PCT on the degree of name concentration in MDB portfolios, we also consider the transition matrix including PCT adjustment reported in Table A3.6 in \cite{RiskControl2023}. The exposure weighted average PD for each MDB portfolio in our data set is reported in Table \ref{tab portfolios summary}.\\

For the MtM approach, we also need the risk-neutral transition rates to derive the market values of the defaultable bonds that constitute the loan portfolio. To convert the historical transition probabilities into risk-neutral transition rates, we follow \cite{Agrawal2004} and \cite{Kealhofer2003} and assume a market Sharpe ratio of $\psi=0.4.$
Risk-neutral probabilities $p^*_s(t,T)$ that an obligor with rating grade $s$ at time~$t$ defaults before time $T$ are then calculated as in the KMV model by
$$
p^*_s(t,T)=\Phi(\Phi^{-1}(p_s(t,T))+\psi \sqrt{T-t} \sqrt{\rho}),
$$
where $\rho$ denotes the asset correlation. Here $p_s(t,T)$ denotes the corresponding historical default probability which can be obtained from the last column of the rating transition matrix in Table \ref{tab transition matrix} for $t=0$ and $T=1.$ Other probabilities can be obtained from the Markovian transition model $p_s(t,T)=p_s(0,T-t)$ for $t\leq T$ (see \cite{HullWhite2000}) and by taking powers of the transition matrix. \\

\begin{figure}[htb]
\begin{center}
       \includegraphics[width=0.8\textwidth]{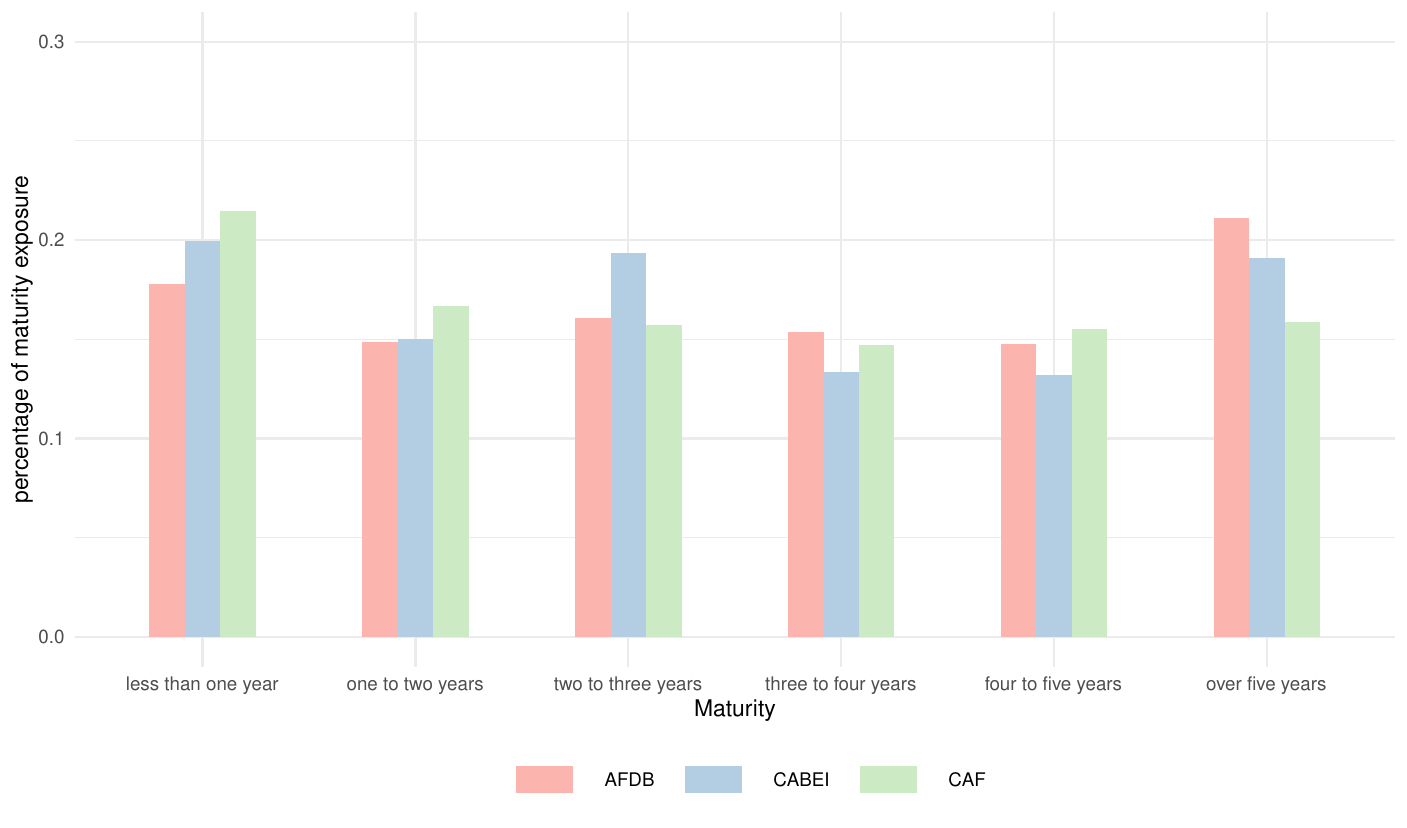} 
\caption{Maturity distribution of MDB portfolios. The figure shows the maturity distribution of the AFDB, CABEI and CAF sovereign loan portfolio as of 2022.} \label{fig Maturity AFDB, CABEI and CAF}
\end{center}
\end{figure}

Information on loan maturities is less explicit in the banks' financial statements, however, a rough distribution of exposures w.r.t. maturities is available for most MDBs in our data sample and is illustrated in Figure \ref{fig Maturity AFDB, CABEI and CAF} for the AFDB, CABEI and CAF loan portfolios as of 2022. Maturity distributions for other MDBs are comparable (see Figure \ref{fig maturity distributions other MDBs} in \ref{app supplementary material}). From these we observe that loan maturities for most MDBs are approximately uniformly distributed. Maturity information is very limited for TDB, EBRD, CDB, however, TDB has a very high exposure expiring in the next 12 months. Table \ref{tab portfolios summary} provides an overview over average maturities for all MDBs in our sample calculated from the information in banks' financial statements.\\

\begin{table}[htb]
    \centering
\scalebox{0.85}{\begin{tabular}{ccccccc}
\toprule
         &  CAF & ADB & AFDB  & IDB & CDB  & CABEI \\
         \hline
         number of borrowers & 16 & 38& 29 &26 &16 &11 \\
         total exposure &28,574 &145,036 & 28,174& 108,520&1,327&9,255\\
         average PD (in \%) &1.46 & 0.18 & 1.46& 0.90&  2.38& 1.46 \\
         average maturity& 5.09& 8.20& 5.62& 8.48&NA& 5.39\\
         \midrule
         & EADB & EBRD & IBRD&  TDB & BOAD&\\
         number of borrowers &4 &38 &78 &21 &8&\\
         total exposure &135& 47,272&229,344 & 6,506&3,868&\\
         average PD & 2.38& 0.9 &0.40 &51.47&2.38&\\
         average maturity& 2.82& NA &7.08&1.64&4.59&\\
         \bottomrule
    \end{tabular}}
    \caption{Summary statistics of MDB portfolios including number of sovereign borrowers (excluding ``regional'' since no ratings for these are available), total exposure (in million USD, converted to USD, if quoted in different currency, based on exchange rate at the date of the financial statement), exposure weighted average PD and exposure weighted average maturity based on information in MDBs financial statements as of 2022.}
    \label{tab portfolios summary}
\end{table}

In our benchmark setting, we consider non-stochastic LGDs of 45\% without PCT. When accounting for PCT, we fix LGDs to 10\%. This is in line with the specification imposed by \cite{S&P2018} for sovereign's borrowing from MDBs with high preferred creditor status. The strong PCT effect is also documented in a recent study by \cite{Fitch2022} who estimate LGDs on MDB sovereign loans between 1.5\% and 12.0\%, indicating that sovereigns' seem to give strong priority to MDBs.\footnote{The PCT effect is measured by comparing LGDs on loans to supranationals and to other creditors.} 
In the case with stochastic LGDs, we set the mean $\ELGD=45\%$ without PCT (resp. 10\% with PCT) and we set the parameter $\nu$ for the calculation of VLGD to $\nu=0.25$ in line with the assumption in the GA formula as applied in \cite{S&P2017,S&P2018}. This implies that volatilities of LGDs are equal to 0.25 without PCT and 0.15 with PCT.\\
%$\nu=0.3$ would implie that volatilities of LGD equal to 0.16 with PCT and 0.27 without PCT which is in line with estimates of haircuts in \cite{CrucesTrebesch2013}.

%Note: For the stochastic LGDs we use the estimates in Cruces and Trebesch (2013) for mean and standard deviations of sovereign haircuts equal to 37.04\% and 27.28\%, resp.. \cite{RiskControl2023} assumes a beta distributed recovery rate with mean $\theta$ and volatility $\lambda\cdot\sqrt{\theta\cdot(1-\theta)}$ and calculates $\lambda=0.2728/\sqrt{0.3704\times (1-0.3704}=0.56$ from the haircut estimates. Moreover, $\theta$ is set to 90\% with PCT and to 55\% without PCT adjustment so that volatilities are 0.168 with PCT and 0.279 without PCT. Our choice is consistent with this approach.\\

In the S\&P methodology, the asset correlation $\rho_n$ for each borrower is determined according to the IRB approach, i.e.\ as a function of the borrower's PD (compare equation (\ref{equ IRB asset correlation})). In our benchmark case, we also consider this specification. In addition, we estimate the asset correlation parameter $\rho$ based on the sovereign default database available from the Bank of Canada and Bank of England\footnote{Compare \url{https://www.bankofcanada.ca/2021/07/staff-analytical-note-2021-15/}.}. The data set is updated on annual basis and covers information on sovereign exposure broken down by country. 
%Default is defined as an event when the borrower cannot repay the debt within the contractual period or when the creditor suffers an economic loss due to a change in the debt (compare \cite{Beers2022}). 
From the database we extract the historical time series of default rates of the IBRD portfolio and use this to estimate the asset correlation parameter $\rho$ based on three different methods: (1) maximum likelihood estimation following \cite{GordyHeitfield2002}, (2) method of moments estimation (see \cite{FreiWunsch2017}) and (3) by estimating the correlation value such that the unexpected loss under the IRB approach matches the quantile of a beta distribution that has been fitted to the empirically observed first two moments of the loss rates (compare \ref{appendix correlation} for details). 
The estimated correlation values range between 29\% to 40\%.

We also estimate asset correlations from equity index returns following the methodology in \cite{RiskControl2023}. We use country equity indices as latent asset return variables and MSCI regional equity indices as single systematic risk factors. Our calculations are based on 35 equity indices of those countries that are borrowing from MDBs in our data set. The resulting regionally averaged estimates for the asset correlation range between 21\% and 51\%. For comparison, Tables A3.2 and A3.4 of \cite{RiskControl2023} imply an asset correlation between 20\% and 68\%. 
Note that we do not use the country specific asset correlations in our analysis, since equity indices are not available for all borrowing countries, actually only for a rather small subset (35 out of 203 countries in total).

%\cite{RiskControl2023} estimates correlations between countries or regions as well as idiosyncratic risk weights based on regional MSCI equity indices and using regional bond yield spreads. Since we do not consider several regional factors but a single systematic risk factor, the asset correlation $\rho_n$ in our model can be derived from their estimated idiosyncratic risk weights $\eta_n$ as $\rho_n=1-\eta_n^2$. Hence, the estimates in Tables A3.2 and A3.4 of \cite{RiskControl2023} imply an asset correlation $\rho$ between 20\% and 68\%. 

%\textcolor{red}{Their estimates also allow to assign asset correlations specific to certain regions (Africa, Europe and Middle East, Asia and Latin America). We might also use these in our empirical study. However, results depend strongly on the estimation method, e.g.\ they obtain high correlations in a certain region when bond spread approach is followed whereas correlations are rather low when equity indices are used. See also Appendix 3 in \cite{RiskControl2023} for the methodology. Table 25 in \cite{S&P2022} classifies countries by regions.}

Thus, we either set the asset correlation according to the IRB approach depending on the borrower's PD or we assign a fixed asset correlation of 0.35 to all borrowers which is in the range for all considered estimation methods above.

Finally, when applying the MtM approach, we assume semi-annual payments, $\delta=0.5$, and we fix the coupon rates to $c=1\%$, since information on spreads is not available for most MDBs in our data set.

The risk-free interest rate is calculated based on the Nelson--Siegel--Svensson method using data on the US yield curve at the end of 2022 (\url{https://www.federalreserve.gov/data/yield-curve-tables/}).

\section{Results}\label{sec results}

In this section, we target two main questions. 
First, we evaluate how relevant single name concentration is for the risk management of MDB portfolios. Secondly, we discuss the appropriateness of the current methodology applied by S\&P to quantify name concentration risk in MDB portfolios and how different parameter specifications affect the GA.
To this end, we calculate and compare the following quantities:
\begin{itemize}
    \item GA MC IRB: exact GA (\ref{equ GA exact actuarial}) calculated by MC simulation within the single factor model (\ref{equ latent asset return}) underpinning the IRB approach, 
    \item GA approx.: full analytical approximation GA (\ref{equ GA 1st order actuarial}) as suggested in \cite{GordyLuetkebohmert2013},
    \item GA simplified: approximate GA in its simplified version (\ref{equ GA 1st order actuarial simplified}) as currently applied in \cite{S&P2018},
     \item GA MtM MC: exact GA (\ref{equ GA exact MtM}) calculated by MC simulation within the MtM default-only and the ratings-based CreditMetrics model,
    \item GA MtM approx.: analytical approximation GA for MtM CreditMetrics model as derived in \cite{GordyMarrone2012}.
\end{itemize}

In line with the IRB approach and the S\&P methodology, we set the confidence level to 99.9\%. We calculate the GAs
\begin{itemize}
    \item with random LGDs, where the volatility parameter $\nu=0.25$,
    \item without randomness in LGDs by setting $\nu=0$.
\end{itemize}

We first compute all GAs for a maturity of one year for all loans and then analyse the impact of different maturities on the GA separately using the average maturity for each borrower in an MDB portfolio according to the values reported in Table \ref{tab portfolios summary}. 
%When applying a (constant) maturity adjustment factor to the exact GA (\ref{equ GA exact actuarial}) in the one-factor actuarial CreditMetrics model similarly to the MA in the approximate GA, this however simply scales up the GAs. Thus, we omit this here and instead consider a constant maturity of one year for all portfolios.\\

Moreover, we evaluate the impact of PCT on the GA. Therefore, we calculate the GAs for both $\LGD=45\%$ and for a reduced $\LGD=10\%$ accounting for a strong PCT effect. Additionally, we study the impact of PCT on GAs through its effect on sovereign default probabilities by applying the PCT-adjusted rating transition matrix as derived in \cite{RiskControl2023}. 
Our main results are summarized in Tables \ref{tab results summary} and \ref{tab results calibrated summary}.

\subsection{Relevance of Name Concentration Risk}

To analyse the relevance of name concentration risk for MDB portfolios, we first focus solely on the exact GA calculations in the single-factor actuarial CreditMetrics model underpinning the Basel IRB approach. 
Our results in Table \ref{tab results summary} clearly document that single name concentration risk is indeed an important risk factor in MDB portfolios. The exact GA calculated by MC simulation in the IRB model ranges between 3\% and 26\% for the benchmark case of non-random $\LGD=45\%$. 
%This clearly demonstrates that name concentration risk is a highly relevant risk factor in MDB loan portfolios.
Stochastic LGDs significantly increase the name concentration risk so that the exact GAs increase roughly by a factor of 1.5 or higher, resulting in values between 4\% and 38\%, when stochastic LGDs with $\nu=0.25$ are considered. This emphasizes the relevance of name concentration risk in MDB portfolios even further. 

As expected, the GAs tend to increase with decreasing number of borrowers (in parentheses in the table). IBRD has the largest portfolio consisting of 78 borrowers and its exact GA calculated by MC simulation in the IRB model is approximately nine times smaller than the GA of the smallest portfolio in our data set (EADB with only 4 borrowers).

By looking at the relative GAs, calculated as the ratio $GA/(\mathcal{K}^* +GA)$ of the GA to total unexpected loss, where $\mathcal{K}^*$ is the unexpected loss in the IRB model, we note that name concentration risk accounts for a very large portion of total unexpected loss. For our benchmark case with $\LGD=45\%$, the GA ratio  for the exact MC GA can be as large as 76\% for the smallest portfolio. When stochastic LGDs are considered, the exact MC GA ratio can be even higher up to 82\%. 
%The corresponding values for the approximate GA and the simplified GA are even slightly higher. 
For comparison, \cite{GordyLuetkebohmert2013} state that the relative GA is between 1\%--8\% for small to large commercial bank portfolios in their data set and can be as large as 40\% of unexpected loss for very small commercial bank portfolios with 250--500 borrowers. This range is also supported by findings in \cite{TarashevZhu2008} and \cite{Heitfieldetal2006} who find that the GA in the actuarial multi-factor CreditMetrics model accounts for 1\%-8\% of total VaR depending on the size of the portfolio. \\
Our results demonstrate that name concentration risk plays a markedly more important role for MDBs whose sovereign loan portfolios are substantially smaller than the very small commercial bank portfolios in the studies mentioned above. 

\begin{landscape}
\begin{table}
\centering
%\begin{adjustbox}{angle=90}
\scalebox{0.60}{
\begin{tabular}{c|c|ccccccccccccc}
\toprule
 $\ELGD=45\%$& & CAF (16)& ADB (38)& AFDB (29) & IDB (26) & CDB (16) & CABEI (11) & EADB (4) & IBRD (78)&  TDB (21) & BOAD (8)& EBRD (37) &\\ \\
 \hline
 \multirow{4}{*}{$\nu=0$} & GA MC IRB &7.29	&4.27&	5.34&	5.48&	8.96&	11.82&	25.19&	2.83&	6.14&	9.94&	5.21\\	
 % 7.29 	4.89 	5.27 	5.97 	8.96 	11.82 	25.19 	2.84 	6.14 	9.94 	5.61
 & GA approx.  &19.30&	12.84&	10.60&	16.23&	15.11&	39.33&	36.90&	4.69&	22.46&	22.00	&9.94	\\
 &GA simplified &19.30&	12.84&	10.60&	16.23&	15.11&	39.33&	36.90&	4.69&	22.46&	22.00	&9.94 \\
%  & GA MC IRB -- GA approx. &7.09&4.52&2.27&5.53&4.18&18.04\\
&rel. GA MC IRB& 46.58	&51.67&	36.57&	46.84&	45.91&	57.32	&75.72&	36.63&	32.33&	48.22&	48.65\\	
%46.58 	47.98 	38.00 	46.84 	45.89 	57.32 	75.72 	36.95 	32.14 	48.22 	47.93
&rel. GA approx. &69.78&	71.76&	55.10&	70.55&	58.84&	81.71&	82.04&	48.48&	63.54&	67.33	&61.56\\
& rel. GA simplified &69.78&	71.76&	55.10&	70.55&	58.84&	81.71&	82.04&	48.48&	63.54&	67.33&	61.56\\
  \hline
 \multirow{4}{*}{$\nu=0.25$}& GA MC IRB&14.33&	7.36	&9.46&	10.83&	15.69&	21.94&	37.83&	4.18&	15.53&	16.74&	9.49\\
 %14.46 	7.73 	8.85 	10.28 	15.62 	21.81 	37.63 	4.48 	15.60 	16.22 	9.42
 &GA approx. &28.78	&19.32&	15.68&	24.40&	21.88&	59.25&	49.97&	6.79&	34.53&	32.93&	14.49\\	
&GA simplified  &25.19	&16.77&	13.84&	21.19&	19.72&	51.35&	48.18&	6.12	&29.33&	28.72&	12.97\\	
&rel. GA MC IRB& 63.37&	59.32&	50.96&	60.24	&58.65&	70.86&	82.23&	46.59&	55.02&	59.98	&60.16\\
% 	63.64 	60.72 	50.97 	60.41 	59.19 	71.03 	82.35 	47.43 	54.43 	60.32 	59.83
&rel. GA approx. &77.49&	79.26&	64.48&	78.27&	67.43&	87.06&	86.09&	57.68&	72.82&	75.51&	70.03	\\
& rel. GA simplified&75.09&	76.83&	61.57&	75.77&	65.11&	85.36&	85.64&	55.13&	69.47&	72.90&	67.64\\
 \midrule\\
  $\ELGD=10\%$& & CAF (16)& ADB (38)& AFDB (29) & IDB (26) & CDB (16) & CABEI (11) & EADB (4) & IBRD (78)&  TDB (21) & BOAD (8)& EBRD (37) &\\ \\
 \hline
 \multirow{4}{*}{$\nu=0$} &GA MC IRB& 1.62&	0.91&	1.17&	1.30&	1.95&	2.65&	5.60&	0.61&	1.34&1.87&	1.30\\	
 & GA approx.  &4.29	&2.85&	2.35&	3.61	&3.36&	8.74	&8.20&	1.04&	4.99&	4.89&	2.21\\
 &GA simplified & 4.29&	2.85&	2.35&	3.61&	3.36&	8.74	&8.20&	1.04&	4.99&	4.89	&2.21\\
%  & GA MC IRB -- GA approx. &7.09&4.52&2.27&5.53&4.18&18.04\\
&rel. GA MC IRB& 46.58&	47.34&	37.82&	47.43&	45.39&	57.32&	75.72&	34.66&	32.26&	47.43&	47.11\\	
&rel. GA approx. &69.78&	71.76&	55.10&	70.55&	58.84&	81.71&	82.04&	48.48&	63.54&	67.33&	61.56\\
& rel. GA simplified  &69.78&	71.76&	55.10&	70.55&	58.84&	81.71&	82.04&	48.48&	63.54&	67.33&	61.56\\
  \hline
 \multirow{4}{*}{$\nu=0.25$}& GA MC IRB&10.46&	7.40&	7.25&	9.62&	11.23&	18.43&	21.53&	3.09&	13.15&	11.40&	6.28\\
 &GA approx. & 19.80&	13.45&	10.66	&16.97&	14.44	&41.34&	29.58&	4.48&	24.74&	22.77&	9.67\\
&GA simplified  &13.94&	9.27&	7.65&	11.72&	10.91&	28.40&	26.65&	3.38&16.22&	15.89	&7.18\\	
&rel. GA MC IRB&84.71&	86.68&	79.09&	86.49&	82.57&	90.34&	92.17&	74.46&	82.45&	83.19&	82.27\\
&rel. GA approx. &91.42&	92.29&	84.75&	91.85&	86.01&	95.48&	94.28&	80.18&	89.62&	90.56&	87.52\\
& rel. GA simplified&88.24&	89.20&	79.95&	88.62&	82.29&	93.56&	93.69&	75.36&	84.99&	87.01&	83.88\\
 \midrule\\
  $\ELGD=10\%$& & CAF (16)& ADB (38)& AFDB (29) & IDB (26) & CDB (16) & CABEI (11) & EADB (4) & IBRD (78)&  TDB (21) & BOAD (8)& EBRD (37) &\\ 
  PCT-adj. PDs&&&&&&&&&&&&&\\
 \hline
 \multirow{4}{*}{$\nu=0$} &GA MC IRB& 1.91&	0.93&	1.12&	1.30&	1.75&	2.93&	3.97&	0.52&	2.00&	2.04&	1.23\\	
 & GA approx.  &3.00&	1.93&	1.67&	2.55&	2.55&	6.03&	7.61&	0.75&	2.98&	3.17&	1.70\\
 &GA simplified &3.00&	1.93&	1.67&	2.55&	2.55&	6.03&	7.61&	0.75&	2.98&	3.17&	1.70\\
%  & GA MC IRB -- GA approx. &7.09&4.52&2.27&5.53&4.18&18.04\\
&rel. GA MC IRB& 57.76&	51.77&	41.78&	52.72&	51.19&	64.89&	78.64&	39.76&	43.43&	47.91&	56.89\\	

&rel. GA approx. &67.90&	69.20&	53.94&	69.31&	60.51&	79.83&	87.59&	48.73&	53.45&	61.84&	63.90\\
& rel. GA simplified  &67.90&	69.20&	53.94&	69.31&	60.51&	79.83&	87.59&	48.73&	53.45&	61.84&	63.90\\
  \hline
 \multirow{4}{*}{$\nu=0.25$}& GA MC IRB&8.44&	6.65	&6.38&	9.13&	8.25	&17.01&	12.22&	2.40&	9.91&	8.84&	5.57\\
 &GA approx. & 11.94&	7.82&	6.59&	10.27&	9.71&	24.32&	26.22&	2.87&	12.36&	12.68&	6.56\\
&GA simplified  &9.76&	6.27&	5.44&	8.28&	8.29	&19.59&	24.72&	2.44&	9.68&	10.31&	5.54\\	
&rel. GA MC IRB&86.44&	88.54&	81.79&	89.00&	82.45&	91.60&	92.17&	75.42&	80.07&	81.72&	85.32\\
&rel. GA approx. &89.38&	90.10&	82.18&	90.10&	85.37&	94.11&	96.05&	78.45&	82.66&	86.63&	87.21\\
& rel. GA simplified&87.30&	87.95&	79.19&	88.01&	83.28&	92.79&	95.82&	75.55&	78.87&	84.04&	85.19\\
 \bottomrule
\end{tabular}
}
%\end{adjustbox}
\caption{GAs for MDB portfolios. The table shows the different GAs (in \% of total EAD) for $q=99.9\%$ for non-random LGDs ($\nu=0$) and with random LGDs for $\nu=0.25$ as well as with and without PCT adjustment. The asset correlation $\rho$ is calculated according to the IRB formula. The relative GA MC IRB refers to the GA MC IRB expressed in \% of unexpected loss, calculated as UL in the IRB model plus GA MC IRB, and analogously for the other relative GAs. Numbers of borrowers are in parentheses.}
\label{tab results summary}
%\end{center}
\end{table}
\end{landscape}

In Table \ref{tab results calibrated summary} we report also the exact GA values for the default-mode and ratings-based MtM CreditMetrics model when assuming a constant maturity of one year (upper panel) or using the mean maturities (lower panel) for each MDB portfolio as reported in Table \ref{tab portfolios summary}. Our results show that the GA values for the default-mode MtM Credit Metrics and the default-model actuarial CreditMetrics model underpinning the IRB approach are relatively close for most MDB portfolios. Comparing the (exact) default-mode and ratings-based MtM GAs indicates that accounting for rating transitions reduces the measure of name concentration risk in most MDB portfolios. Further, by looking at the corresponding GA values for the average maturity case, we see that allowing for rating transitions and correctly adjusting for maturities, mostly leads to a reduction of the GA in the non-random LGD case. In contrast, when LGDs are random, the GAs tend to increase when average maturities are used instead of one year maturities and when rating transitions are taken into account instead of the default-only case. We discuss this effect further in Section \ref{subsec parameter choice}. Our results indicate that the variance of the LGD variable has to be very carefully chosen since it has a strong impact on the GA values. Overall, the GA values remain at a very high level which underlines the relevance of name concentration risk in MDB portfolios. \\

\subsection{Appropriateness of S\&P Approach}

Next, we analyse the appropriateness of the leading methodology as implemented by S\&P for the measurement of name concentration in MDB portfolios. To this end, we consider both the full analytical approximation GA and the simplified GA version of \cite{GordyLuetkebohmert2013} and compare them with the exact GA based on MC simulation in the IRB model. First, we note from Table \ref{tab results summary} that there is a considerable gap between the exact GA and the analytical approximation GA, especially for the very small portfolios. This difference between the GA MC IRB and the GA approx. can be as high as 28 percentage points for constant LGDs and up to 37 percentage points in the random LGD case (compare CABEI portfolio). The gap is much lower (in absolute terms) for the largest portfolio (IBRD) in our data set than for the other MDBs, documenting that the analytic GA indeed becomes less accurate when portfolios become very small. However, even for the IBRD portfolio the approximation GA overestimates the exact GA by about 65\% in the benchmark case and 62\% in the stochastic LGD case. The largest overestimation is visible in the TDB portfolio with a 266\% increase from the exact to the approximate GA in the benchmark case.

%When comparing results in Table \ref{tab results 99.9 LGD 45} with those in Table \ref{tab results 99 LGD 45}, we see that the analytic approximation is more accurate for the lower quantile $q=99\%$ than for $q=99.9\%.$ For the extremely small portfolio of EADB, the exact MC GA is even larger than the analytic GA for $\nu=0.$ \\

Moreover, our results for the random LGD case show that the simplified GA is significantly lower than the full analytic approximation (GA approx.) for most MDBs which is due to the fact that the simplified GA ignores higher order terms in PDs.\footnote{Note that the simplified GA coincides with the approximate GA in the case of non-random LGDs as can be easily seen from the expression in (\ref{equ GA 1st order actuarial}).} These, however, have a non-negligible effect since the average PDs for the MDB portfolios in our data set can be relatively large as shown in Table \ref{tab portfolios summary}. 
%The effect is more pronounced for $q=99.9\%$ than for $q=99\%$. 
While this reduces the gap between the approximate GA and the exact MC GA, it lacks theoretical justification given the high average default probabilities of the borrowers in MDB portfolios.   \\

The huge gap between the exact and the approximate GA can to some extend be explained by an inappropriate calibration of the underlying model parameters.
The GA formula (\ref{equ GA 1st order actuarial}) is based on the CreditRisk$^+$ model, which assumes a Gamma distributed risk factor with mean 1 and variance $1/\xi$. In \cite{S&P2018} this parameter is set to $\xi=0.25.$ While this is in line with \cite{GordyLuetkebohmert2013}, the authors calibrated $\xi$ by matching the approximate GA for a representative commercial bank portfolio to the exact GA implied by the MtM CreditMetrics model. Hence, this parameter might not be appropriate for typical MDB portfolios. To investigate this further, we calculate on the one hand the exact GA in the ratings-based CreditMetrics model using the average maturity for each MDB portfolio (compare Table \ref{tab portfolios summary}) and a constant coupon rate of 1\% and on the other hand the approximate GA using the maturity adjustment $\MA$ corresponding to the average maturity of each portfolio and determine the free parameter $\xi$ such that both quantities agree. The resulting values are reported in Table \ref{tab calibrated xi values} and show a substantial variation across MDBs.

\begin{table}
    \centering
\scalebox{0.7}{
    \begin{tabular}{c|ccccccccccc}
    \toprule
       & CAF  &   ADB   &  AFDB  &  IDB  &   CDB   &  CABEI  & EADB  &  IBRD  &  TDB   &  BOAD  &  EBRD\\
       \hline
        $\xi$ & 0.11 	&0.09 	&0.57 	&0.12 	&0.42 	&0.02 	&0.05 	&1.08 	&0.06 	&0.12 	&0.21\\
        \bottomrule
    \end{tabular}}
    \caption{Calibrated values for variance parameter $\xi$}
    \label{tab calibrated xi values}
\end{table}

For regulatory purpose, a unique value for $\xi$ would, of course, be preferable. For this reason and to avoid overfitting, we set $\xi$ in the following analysis such that the mean squared error between the exact and the approximate GA is minimized across all eleven portfolios. This results in a value $\xi=0.063$, which is considerably lower than the original choice of $0.25$.
Table \ref{tab results calibrated summary} reports the corresponding exact and approximate GA values for the MDB portfolios using this choice of $\xi$. In addition to the results for the portfolios with average maturities and for better comparability with the results in Table \ref{tab results summary}, we also report the GA values for one year maturity across all portfolios. We observe that the gap between the approx. GA and the GA MC IRB is still significant since the parameter $\xi$ was calibrated by minimizing the mean squared error over all eleven portfolios. For some portfolios the approx. GA is now relatively close to the MC GA in the ratings-based MtM CreditMetrics model, e.g.\ for ADB, EADB and TDB for $\nu=0.25$ when average maturities are used, which is not surprising because the values for $\xi$ when calibrated for these individual portfolios are very close to the value $0.063$ used in the GA calculations reported in Table \ref{tab results calibrated summary}. For others, however, the approx. GA can substantially deviate; compare e.g.\ AFDB, CDB, CABEI, IBRD or EBRD for the same setting. In particular, for the non-random LGD case and the one year maturity case, the approximation does not perform very well since $\xi$ was calibrated for the random LGD case with mean maturities. Overall, we can state that calibrating $\xi$ can reduce the gap between the approx. GA and the exact GA, but there is no parameter that is equally well suited for all MDB portfolios compromising the use of this approach for credit rating agencies' assessment of MDBs' capital adequacy.

\subsection{Impact of Model Parameters}\label{subsec parameter choice}

\begin{figure}
 \subfigure[GA MC IRB (in \% of total EAD)]{
         \includegraphics[width=0.5\textwidth]{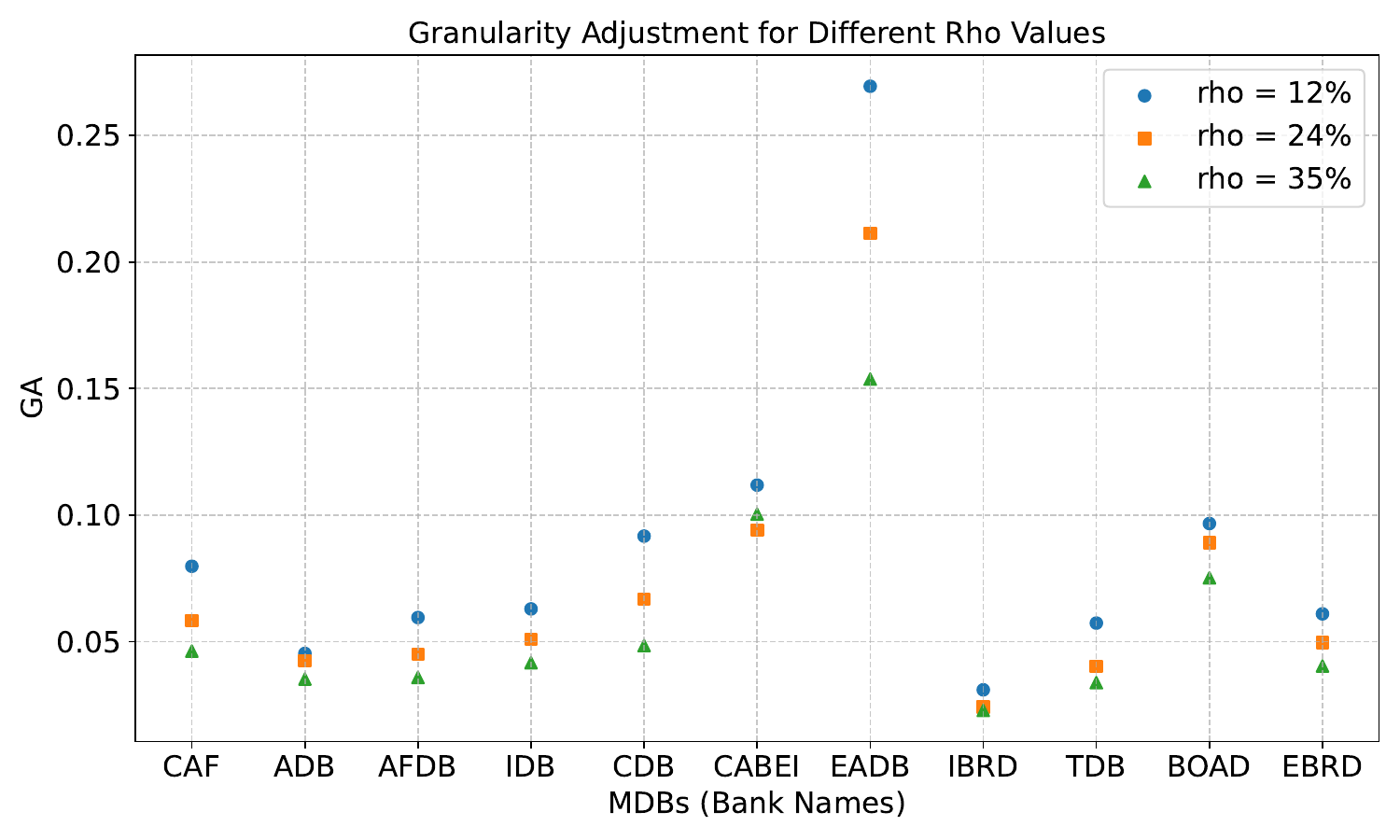}}
\subfigure[GA MC IRB (in \% of total UL)]{
         \includegraphics[width=0.5\textwidth]{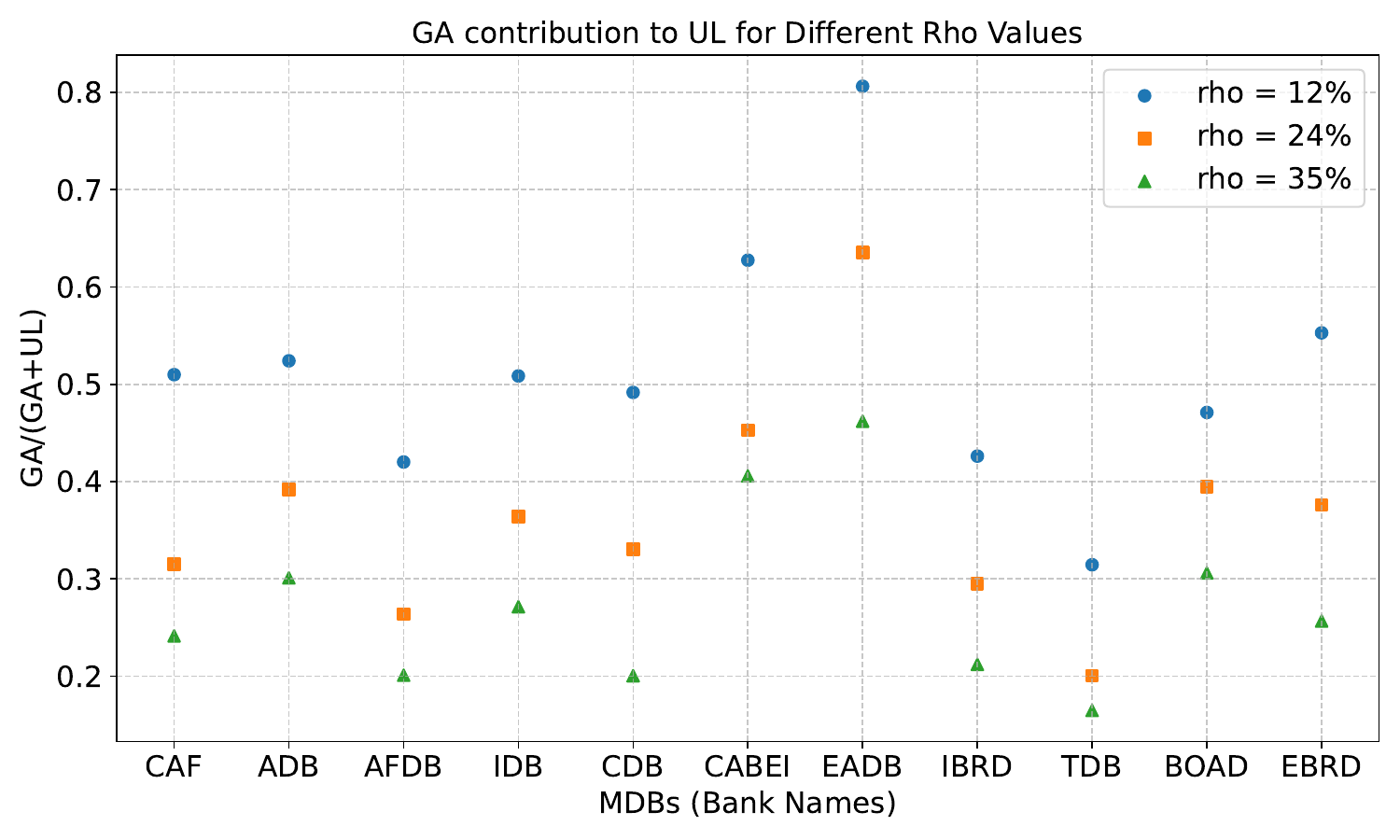}}
    \caption{GAs for different asset correlations. The figure shows the GA MC IRB in \% of total EAD (left panel) and in \% of total UL (right panel) in the benchmark case with $\ELGD=45\%$ and $q=99.9\%$ calculated for constant asset correlation $\rho=12\%,\, 24\%$ and 35\%. LGDs are non-random LGDs ($\nu=0$).}
    \label{fig:results_rho}
\end{figure}

Next, we discuss the choice of the model parameters and their influence on the GA.
Therefore, we first consider the effect of the asset correlation parameter $\rho$ on the size of the GA. The methodology in \cite{S&P2018} relies on the IRB asset correlation which by construction is between 12\% and 24\%. As mentioned earlier, a recent study of \cite{RiskControl2023} as well as our own estimations based on historical default rate time series indicate a higher asset correlation around 35\% (compare Section \ref{sec data}). Our results in Figure~\ref{fig:results_rho} show that the MC GA in the single factor actuarial CreditMetrics model with constant asset correlation substantially decreases for most MDB portfolios when raising the asset correlation from a constant level of 12\% to 24\% or 35\%. The intuition is that larger asset correlation reduces idiosyncratic risk and hence lowers name concentration risk. This is also in line with findings in \cite{GordyMarrone2012}. Thus, by adhering to the IRB asset correlation formula, the approach in \cite{S&P2018} rather overestimates the name concentration risk in MDB portfolios since actual correlations for sovereigns seem to be higher implying lower GAs.
%However, in the MC GA in the IRB model, asset correlations vary across borrowers giving rise to additional name concentration risk. This partly offsets the effect of decreasing idiosyncratic risk when $\rho$ rises.\\

\begin{landscape}
\begin{table}
\centering
%\begin{adjustbox}{angle=90}
\scalebox{0.60}{
\begin{tabular}{c|c|ccccccccccccc}
\toprule
 Maturity 1 year& & CAF (16)& ADB (38)& AFDB (29) & IDB (26) & CDB (16) & CABEI (11) & EADB (4) & IBRD (78)&  TDB (21) & BOAD (8)& EBRD (37) &\\ \\
 \hline
 \multirow{7}{*}{$\nu=0$} & GA MC IRB & 	5.69 	&4.45 	&4.52 	&4.98 	&6.01 	&11.12 	&16.35 	&2.76 	&3.65 	&7.29 	&4.69\\	
 & GA approx.  &14.57 	&9.72 	&7.98 	&12.28 	&11.32 	&29.77 	&27.13 	&3.51 	&17.10 	&16.62 	&7.45	\\
 &GA simplified &14.57 	&9.72 	&7.98 	&12.28 	&11.32 	&29.77 	&27.13 	&3.51 	&17.10 	&16.62 	&7.45	\\
  &GA def. MtM approx. & 	13.56 	&7.40 	&7.16 	&9.79 	&12.58 	&23.98 	&31.85 	&3.18 	&14.53 	&14.88 	&6.69\\
  &GA rat. MtM approx. &11.46 	&5.92 	&6.15 	&7.86 	&11.45 	&19.26 	&31.41 	&2.89 	&11.09 	&12.69 	&5.92\\
 &GA def. MtM MC  &8.32 	&4.65 	&5.56 	&6.27 	&9.63 	&12.21 	&24.39 	&2.93 	&7.41 	&9.53 	&5.77	\\
 &GA rat. MtM MC  &7.53 	&4.33 	&5.15 	&5.92 	&9.02 	&11.64 	&24.06 	&2.68 	&6.40 	&9.56 	&5.58	\\
 \hline
 \multirow{7}{*}{$\nu=0.25$}& GA MC IRB&10.67 	&6.52 	&7.03 	&8.21 	&10.36 	&17.15 	&26.67 	&3.94 	&11.58 	&12.85 	&6.92\\
 &GA approx. &21.64 	&14.56 	&11.76 	&18.39 	&16.32 	&44.67 	&36.59 	&5.07 	&26.17 	&24.78 	&10.84\\	
&GA simplified  &19.02 	&12.68 	&10.42 	&16.03 	&14.77 	&38.86 	&35.42 	&4.58 	&22.33 	&21.70 	&9.73\\	
&GA def. MtM approx. &  	23.87 	& 13.87 	& 12.36 	& 17.92 	& 19.64 	& 44.54 	& 42.86 	& 5.19 	& 31.25 	& 27.79 	& 10.91\\
&GA rat. MtM approx.& 23.87 	&13.87 	&12.36 	&17.92 	&19.64 	&44.54 	&42.86 	&5.19 	&31.25 	&27.79 	&10.91\\
&GA def. MtM MC  &11.84 	&7.16 	&8.76 	&9.11 	&12.49 	&16.29 	&24.51 	&4.63 	&16.41 	&15.49 	&7.54\\
 &GA rat. MtM MC  &18.86 	&10.40 	&12.68 	&13.47 	&18.77 	&26.02 	&36.40 	&6.39 	&25.93 	&22.69 	&11.37\\
 \midrule
  Mean Maturities& & CAF (16)& ADB (38)& AFDB (29) & IDB (26) & CDB (16) & CABEI (11) & EADB (4) & IBRD (78)&  TDB (21) & BOAD (8)& EBRD (37) &\\ \\
 \hline
 \multirow{7}{*}{$\nu=0$} &GA MC IRB &5.68 	 &4.41 	 &4.55 	 &5.01 	 &6.12 	 &11.22 	 &16.74 	 &2.83 	 &3.59 	 &7.34 	 &4.72\\
 & GA approx.  &12.40 	&7.62 	&7.03 	&9.18 	&10.03 	&24.49 	&26.34 	&3.18 	&16.75 	&15.09 	&6.22\\
 &GA simplified   &12.40 	&7.62 	&7.03 	&9.18 	&10.03 	&24.49 	&26.34 	&3.18 	&16.75 	&15.09 	&6.22\\
 &GA def. MtM approx.  & 	10.80 	&4.98 	&5.73 	&6.73 	&11.58 	&16.71 	&32.08 	&2.78 	&14.53 	&11.78 	&5.66\\
 &GA rat. MtM approx.  &5.71 	&3.12 	&4.61 	&3.89 	&6.71 	&9.81 	&26.39 	&2.09 	&11.09 	&8.74 	&3.56\\
 &GA def. MtM MC  &7.84 	&4.63 	&5.47 	&6.16 	&9.73 	&12.14 	&25.17 	&3.08 	&7.30 	&9.76 	&6.16\\
 &GA rat. MtM MC  &4.73 	&3.45 	&4.78 	&3.98 	&5.47 	&8.63 	&20.41 	&2.06 	&6.51 	&7.61 	&3.39\\
  \hline
 \multirow{7}{*}{$\nu=0.25$}& GA MC IRB &10.67 	&6.55 	&7.08 	&8.13 	&10.26 	&17.14 	&26.68 	&3.86 	&11.43 	&12.89 	&6.94	\\
 &GA approx.  &18.40 	&11.19 	&10.28 	&13.63 	&14.53 	&36.65 	&35.74 	&4.52 	&25.64 	&22.43 	&9.00\\
&GA simplified   & 	16.18 	&9.95 	&9.17 	&11.99 	&13.10 	&31.98 	&34.39 	&4.16 	&21.87 	&19.70 	&8.12	\\
&GA def. MtM approx.  &38.74 	&24.35 	&18.65 	&31.06 	&26.57 	&72.46 	&42.98 	&7.46 	&31.25 	&48.26 	&14.38\\
 &GA rat. MtM approx.  &24.18 	&13.23 	&13.20 	&17.52 	&20.67 	&46.92 	&36.48 	&4.62 	&31.25 	&32.66 	&9.43\\
  &GA def. MtM MC  &16.44 	&10.67 	&13.17 	&13.66 	&17.95 	&21.47 	&25.46 	&7.46 	&16.34 	&21.61 	&10.71\\
 &GA rat. MtM MC  &21.04 	&12.30 	&16.03 	&15.94 	&21.83 	&27.85 	&34.74 	&8.00 	&25.71 	&26.30 	&11.93\\
 \bottomrule
\end{tabular}
}
\caption{GAs for MDB portfolios. The table shows the different GAs (in \% of total EAD) for $q=99.9\%$ for maturity constant equal to 1 year and using the average maturities for each portfolio as reported in Table \ref{tab portfolios summary}. For the approximate GA the calibrated parameter $\xi=0.063$ is applied. LGDs are either non-random ($\nu=0$) or random with $\nu=0.25$ and $\ELGD=45\%$. The asset correlation $\rho$ is calculated according to the IRB formula. Numbers of borrowers are in parentheses.}\label{tab results calibrated summary}
%\end{center}
\end{table}
\end{landscape}

\begin{figure}
\subfigure[Exact MtM GA for random LGD]{
         \includegraphics[width=0.5\textwidth]{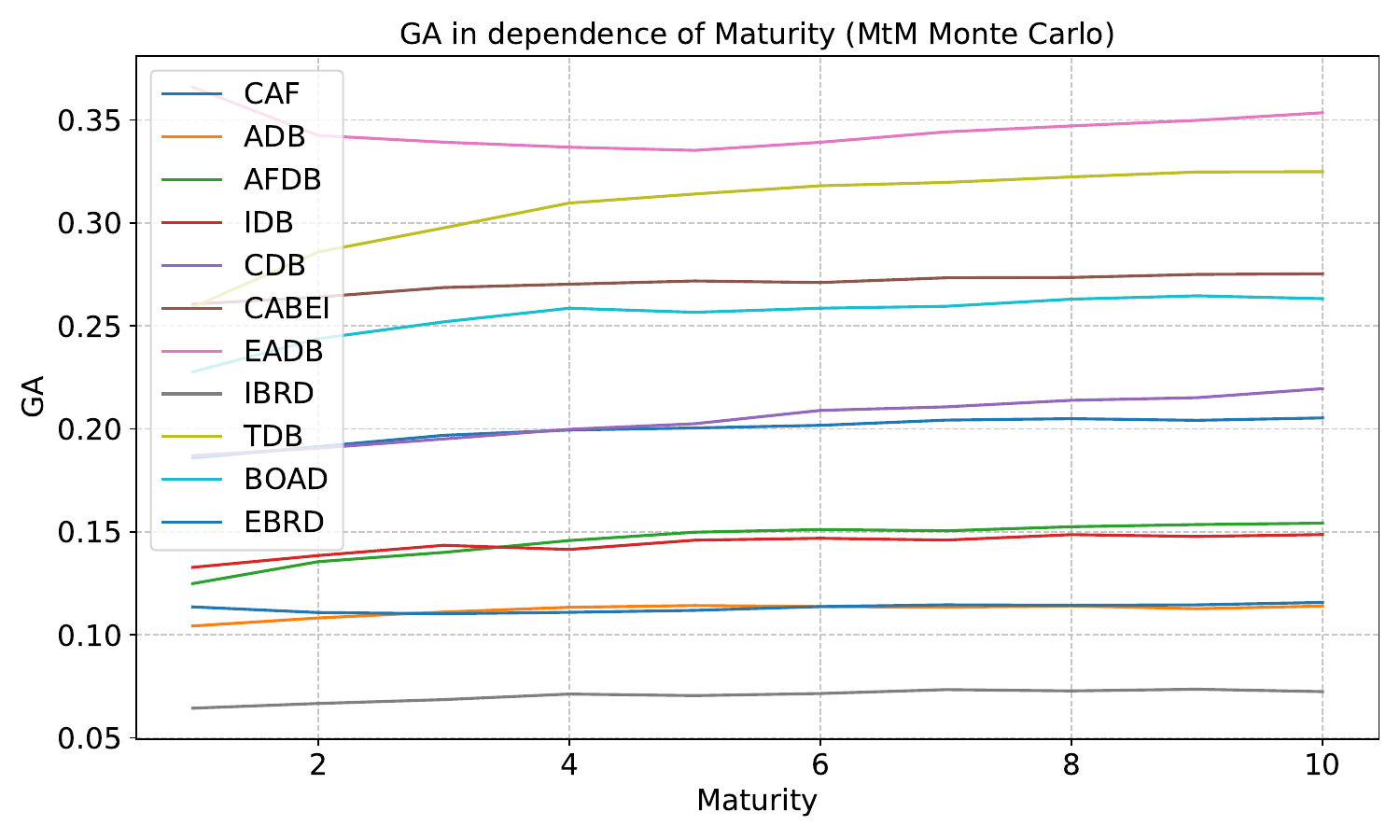}} 
%\subfigure[Approx. MtM GA (in \% of total UL)]{
 %        \includegraphics[width=0.5\textwidth]%{Figure/mdbs_dependence_on_maturity_mtm_approx.pdf}}
 \subfigure[Exact MtM GA for non-random LGD]{
         \includegraphics[width=0.5\textwidth]{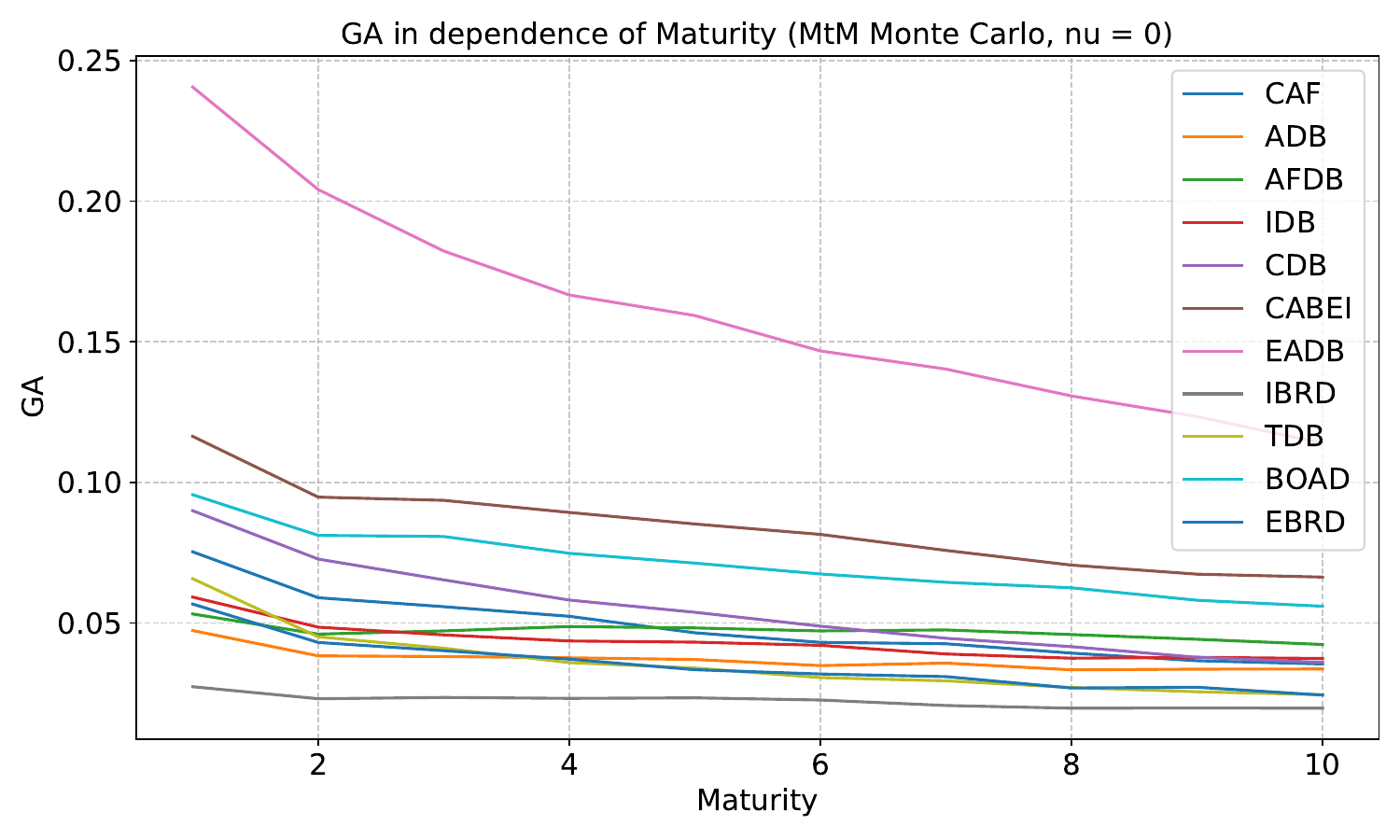}} 
\subfigure[Approx. GA for random LGD]{
         \includegraphics[width=0.5\textwidth]{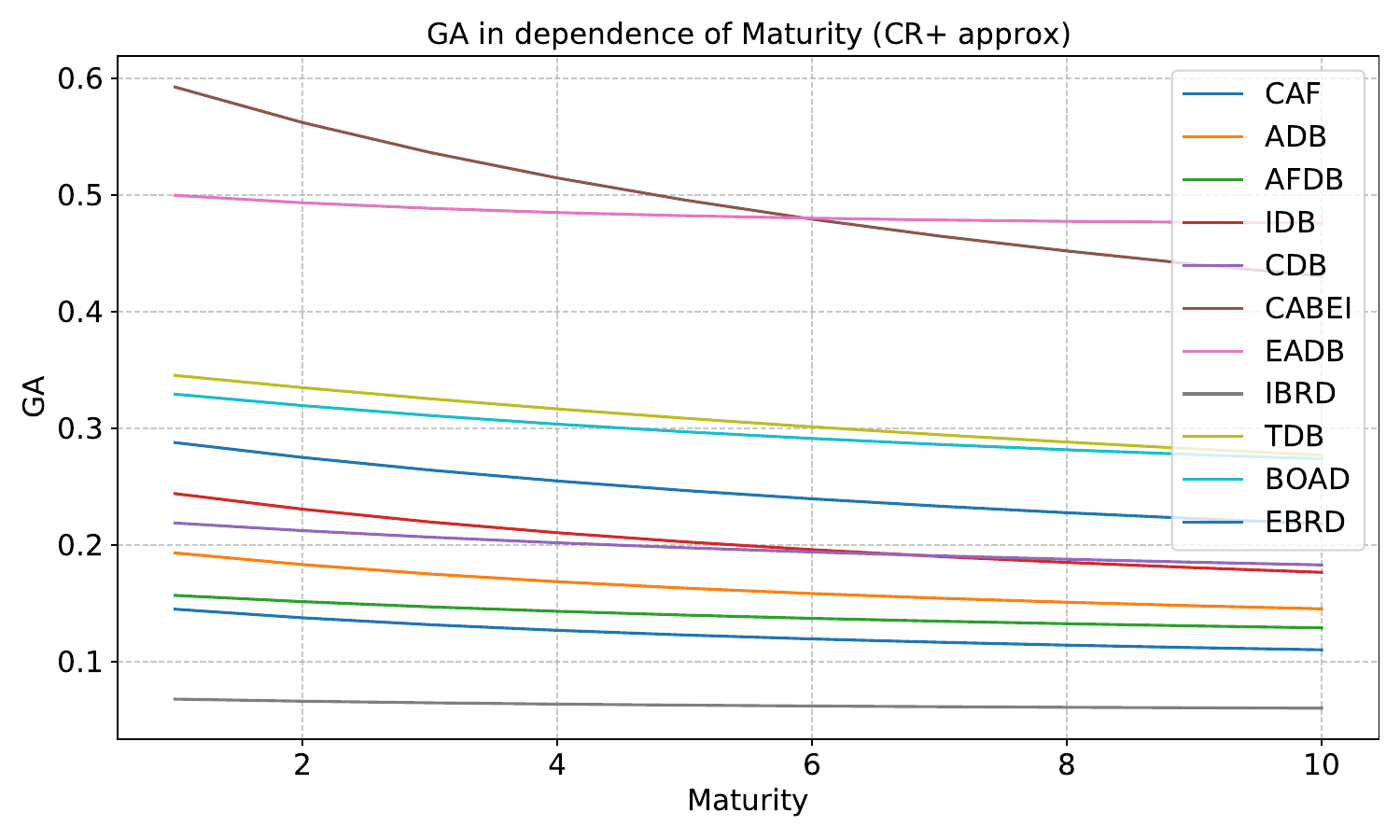}}  
\subfigure[Approx. GA for non-random LGD]{
         \includegraphics[width=0.5\textwidth]{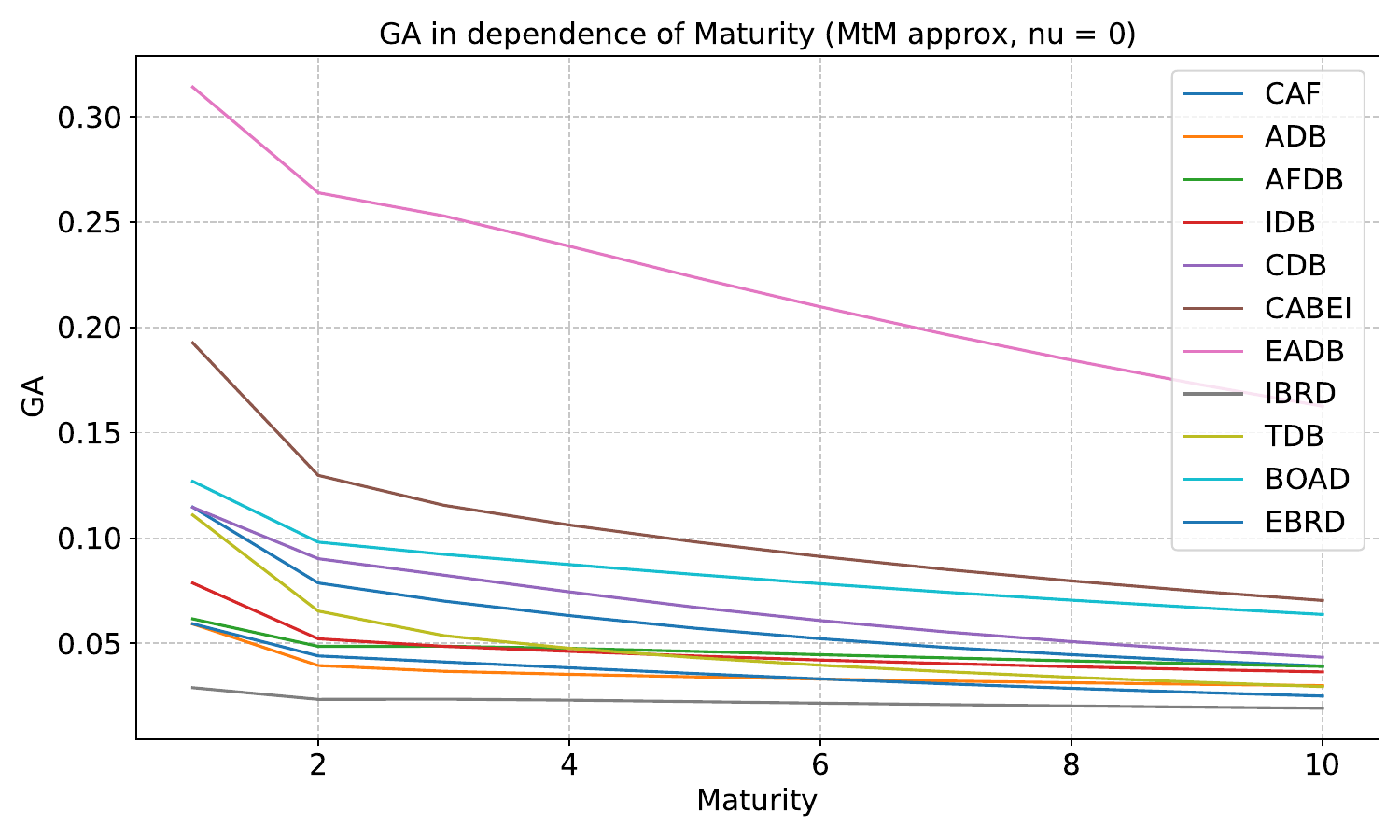}} \caption{GAs for $q=99.9\%$ for different maturities. The figure shows the exact GA (in ˙\% of total EAD) 
         %and the approximate GA (Panel b) 
         in the ratings-based MtM CreditMetrics model for random (Panel a) and non-random LGD (Panel b) as well as the approximate GA for random (Panel c) and non-random LGD (Panel d) in the CreditRisk$^+$ setting. $\ELGD=45\%$ and $\nu=0.25\%$ when LGDs are random. The asset correlation is calculated according to the IRB formula.}
    \label{fig maturity effect on GAs}
\end{figure}

The results in Table \ref{tab results summary} are all based on the assumption that loans have a fixed maturity of one year. Most of the loans in the realistic MDB portfolios, of course, have a longer maturity. 
To study the impact of varying maturities on the measures of name concentration risk, we calculate the exact GA in the ratings-based MtM CreditMetrics model as well as the approximate GA in the CreditRisk$^+$ setting for different maturities. The results are illustrated in Figure \ref{fig maturity effect on GAs}. 
When LGDs are non-random, the exact GA in the MtM model decreases with increasing maturity across all portfolios in our data set (see Panel b).
The rationale for this is as follows. If bond maturities increase, returns are more sensitive to rating transitions and default risk. This affects both the VaR of the portfolio loss rate and the conditional expected loss, albeit in different ways. The conditional loss is more sensitive to rating transitions since it is conditional on an adverse systematic draw which has a negative impact on all loans in the portfolio resulting in several downgrades. In contrast, the VaR is more sensitive to rising default risk since large losses are mainly driven by adverse idiosyncratic draws. When LGDs are non-random, the former effect is dominating so that the GA as the difference between the VaR and the conditional expected loss decreases with increasing maturity. Introducing randomness to the LGDs has no impact on the conditional loss since the latter only depends on the $\ELGD$. However, the VaR increases due to the risk associated with large realizations of the LGD variable. With the variance parameter $\nu$ being large enough this effect leads to an increasing GA with rising maturity (see Panel a).  Overall, the effect is more pronounced for portfolios of low average rating (see, e.g., TDB) than for portfolios of higher average rating (see, e.g., EBRD). This also supports our findings in Table \ref{tab results calibrated summary}.

The approximate GA (\ref{equ GA 1st order actuarial}) implicitly adjusts for different maturities through the MA factor, which equals one when the maturity is one year and is larger one for longer maturities. 
Thus, the approx.\ GA in the CreditRisk$^+$ setting only depends on maturity through the maturity-adjusted UL capital requirements $\mathcal{K}_n$. While these increase with increasing maturity, they are scaled afterwards by the squared exposure shares $a_n^2$ and are then divided by the total UL capital $\mathcal{K}^*$ which is the sum of UL capital requirements $\mathcal{K}_n$ weighted by the exposure shares $a_n$. Thus, the MA factor enters both the numerator and the denominator but is scaled by a smaller factor in the numerator than in the denominator so that in total the approx. GA slightly decreases with increasing maturity (compare Fig. \ref{fig maturity effect on GAs}, Panels c and d). Hence, when compared to the exact GA in Panel (a) when LGDs are random, the approximate GA does not correctly adjust for the impact of rising maturities in concentrated portfolios of low average credit quality as those typically held by MDBs.\\

%\begin{figure}
%\subfigure[Concentrated TDB Portfolio]{
%         \includegraphics[width=0.5\textwidth]{Figure/mdbs_dependence_on_maturities_tdb.pdf}} 
%\subfigure[Homogeneous Portfolio ($N=100$ and B- rating)]{
%         \includegraphics[width=0.5\textwidth]
%         {Figure/mdbs_dependence_on_maturities_homo_B-.pdf}} 
%\subfigure[Homogeneous Portfolio ($N=100$ and BB- rating)]{
%        \includegraphics[width=0.5\textwidth]
%         {Figure/mdbs_dependence_on_maturities_homo_BB-.pdf}} %to be replaced by new figure
%\subfigure[Homogeneous Portfolio ($N=100$ and AA rating)]{
%         \includegraphics[width=0.5\textwidth]
%         {Figure/mdbs_dependence_on_maturities_homo_AA.pdf}} %to be replaced by new figure
%         \caption{Total VaR and asymptotic VaR for $q=99.9\%$ for different maturities. The figure shows the VaR and the conditional loss given the quantile $\alpha_q(X)$ of the risk factor (asymptotic VaR) in the ratings-based MtM CreditMetrics model for the concentrated TDB portfolio (Panel a) and for a homogeneous portfolio with $N=100$ and B- rating (Panel b), BB- rating (Panel c) as well as AA rating (Panel d). GAs and VaRs are calculated using MC simulation and the analytical approximation formula. The LGD is random with $\ELGD=45\%$ and $\nu=0.25\%$ and the asset correlation is calculated according to the IRB formula. Coupons and riskfree rate are set to 0 to eliminate the effect of discounting and additional cash flows.}
%    \label{fig maturity asymptotic VaR}
%\end{figure}

\begin{figure}
\centering
         \includegraphics[width=0.8\textwidth]{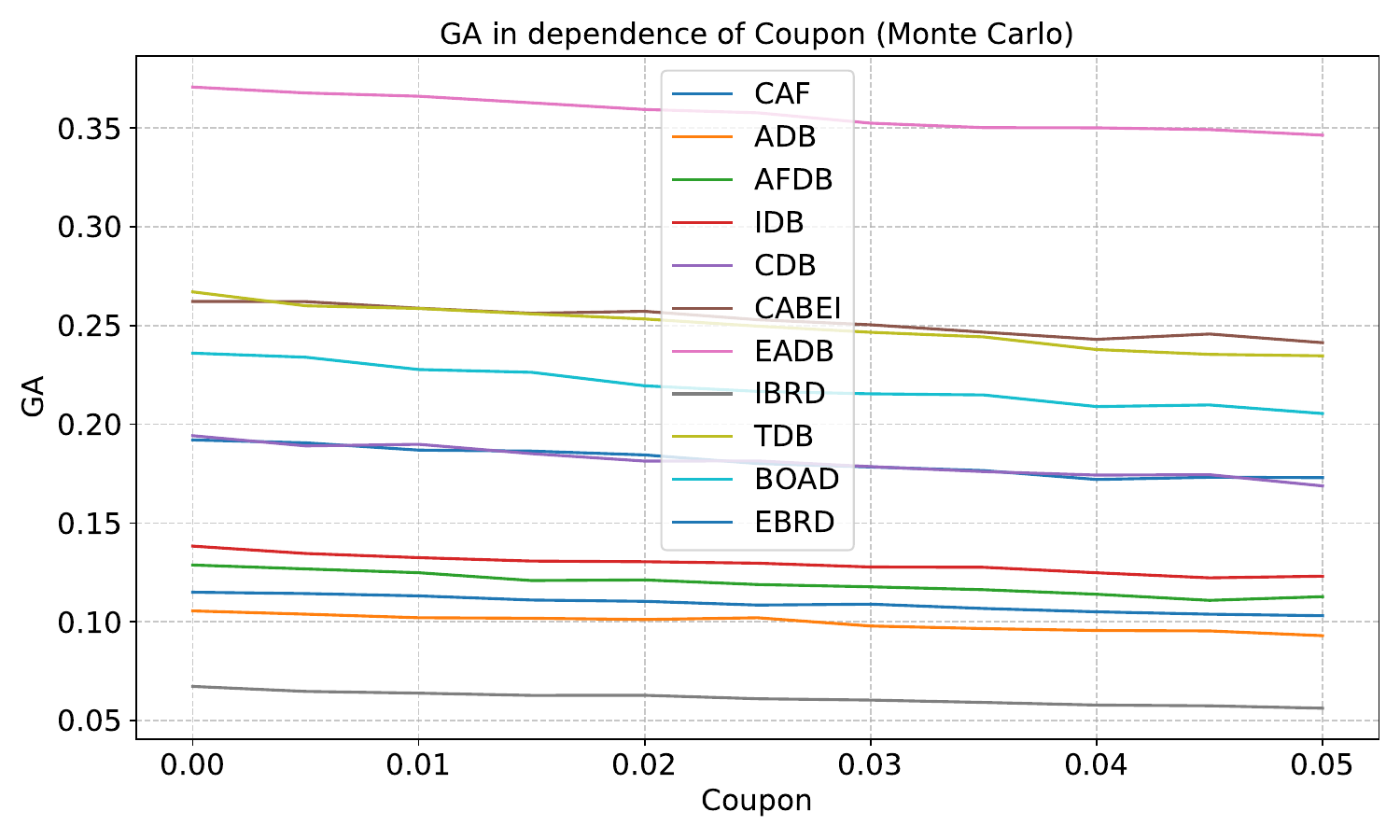}
\caption{GAs for $q=99.9\%$ for different coupon rates. The figure shows the dependence of the exact GA  
         %and the approximate GA (Panel b) 
         in the ratings-based MtM CreditMetrics model (in \% of total EAD) on the coupon rate for the eleven MDBs in our data set. The LGD is random with $\ELGD=45\%$ and $\nu=0.25\%$ and the asset correlation is calculated according to the IRB formula. Maturities are fixed to the average maturities for each portfolio as reported in Table \ref{tab portfolios summary}.}
    \label{fig GA sensitivity to coupon rate}
\end{figure}

Since the information on spreads on sovereign loans is very limited or even not available at all from the financial statements of the MDBs in our data set, we assume a fixed coupon rate of 1\% in our empirical study. We analyse in Figure \ref{fig GA sensitivity to coupon rate} how sensitive our results are to this choice. The figure indicates that the GA decreases with increasing coupon rate across all portfolios. The rationale for this effect is that higher coupon rates decrease the duration of the coupon bonds representing the loans. Hence, the impact of increasing coupons on the GA is opposite to the effect of increasing maturities, resulting in a decreasing GA.\\
%The asymptotic VaR $\mathbb{E}[L|\alpha_q(X)]$ is less affected by changing coupon rates since we already condition on a bad systematic state where a default is quite likely. The true VaR $\alpha_q(L)$, however, decreases when the duration decreases because payments are then less sensitive to rating migrations further in the future. 

Next, we investigate whether name concentration risk becomes less important when MDBs' preferred credit status is taken into account. First, we observe that the GA scales with LGD when $\nu=0$ due to the positive homogeneity of the VaR. Thus, reducing LGD from 45\% to 10\% (middle rows in Table \ref{tab results summary}) scales the GAs down by a factor 10/45. When LGDs are stochastic with $\nu=0.25$, switching from $\LGD=10\%$ to 45\% increases the GA by up to 76\%. 

While PCT lowers LGD rates, it also reduces sovereign PDs. In the last rows of Table \ref{tab results summary}, we calculate GAs for $\LGD=10\%$ and using the PCT-adjusted default rates as obtained in \cite{RiskControl2023}. 
Our results show that the GA in percentage of total EAD decreases to 0.5\% -- 4.0\% in the constant LGD case and 2.4\% -- 17.0\% in the random LGD case, although this effect is mostly due to the PCT-adjusted LGDs. We further note that the relative GAs (as fraction of total UL) can be even larger than in the situation without PCT adjustments, taking values up to 92\% of total unexpected loss. This is due to the fact that lower LGDs and PDs decrease the unexpected loss and hence raise the relative effect of name concentration risk.
Overall, this clearly points out the important role of single name concentration risk as a major risk factor in MDB portfolios even when considering PCT-adjusted inputs.

\cite{S&P2018} accounts for PCT by varying LGDs between 10\% and 45\% depending on the strength of the PCT effect for institutions and using PCT-adjusted default probabilities. Our numerical results indicate that the GA is highly sensitive to the way PCT effects are considered, and hence particular attention has to be paid to correctly measure PCT effects on LGDs and PDs for individual institutions. The study of \cite{RiskControl2023} provides a suitable approach for this.

Finally, we point out that one of the most important techniques used by MDBs to reduce the penalty for single name concentration risk in recent years is Exposure Exchange Agreements (EEAs) with other MDBs. These allow to exchange exposures to sovereign borrowers in which an MDB is concentrated against exposures to sovereigns in which the MDB has no or low exposure. %same amount of credit exposure is exchanged, the EEA seller compensates the EEA buyer at an agreed upon rate, MDBs need to keep a minimum of 50\% of total exposure to each country included in an EEA 
As we do not have publicly available data on individual EEAs for MDBs, we do not explicitly quantify the impact of these agreements on the GAs. However, although EEAs may reduce name concentration risk, they may also have other negative impacts and policy implications that need to be considered.

\section{Conclusion}\label{sec conclusion}

In this paper, we construct realistic MDB portfolios based on publicly available data and investigate the magnitude of the exposure to single name concentration risk in MDB sovereign loan portfolios as well as the accuracy of the approximate GA as currently applied by S\&P. Our results demonstrate that MDB portfolios are substantially exposed to single name concentration risk with GAs accounting for up to 92\% of total unexpected loss. The accurate consideration of this risk source is therefore of utmost importance for the rating agencies' assessment of MDBs' capital adequacy.

However, we show that the analytic approximation GA applied by rating agencies loses accuracy when portfolios consist of only very few borrowers. When applied to MDB portfolios, the approach is overly conservative as it can lead to an overestimation of up to 266\% of the exact GA for the MDB portfolios in our data set. Moreover, we point out the importance of a correct specification of the model parameters. The input parameters to the approximate GA formula have been calibrated to commercial bank portfolios and our results show that these are not always appropriate when applied to MDB sovereign loan portfolios. 

Taken together the approximate GA as currently implemented by S\&P may lead to a significant overestimation of name concentrations in MDB portfolios. This may result in a too conservative assessment of MDBs' creditworthiness making it more difficult for MDBs to raise funding on internal capital markets. This in turn may severely restrict MDBs' lending headroom which has an impact on the achievement of their development goals.

\section*{Acknowledgment}

This project is funded by the MDB Challenge Fund. The MDB Challenge Fund is administered by New Venture Fund and supported by grants from the Bill \& Melinda Gates Foundation, Open Society Foundations and the Rockefeller Foundation. Financial support is gratefully acknowledged. Further, we thank the participants of the workshop on name concentration risk in MDB portfolios for the insightful discussion and valuable remarks. We are particularly thankful to Chris Humphrey and Maura Cravero for various helpful comments.

%\section*{Declarations of Interest}
%The authors have no relevant financial or non-financial interests to disclose.

%\section*{Availability of data and material}
%Data is available through Bloomberg. Code is available on GitHub.

\bibliographystyle{agsm}
	\bibliography{Literature}

\appendix

\section{Asset Correlation}\label{appendix correlation}

 The approach by \cite{Fitch2008} is based on the idea that the asset correlation should be reflected in the realized volatility of portfolio losses over time. Hence, by fitting the mean and standard deviation of historical default data to some specific distribution and matching the expected and unexpected losses calculated based on the resulting loss distribution to the EL and UL under the Basel regulatory framework, one can infer the asset correlation value. More specifically, based on a robust time series of observed losses for a certain asset class, the first two moments of the loss rates are calculated. The mean corresponds to the expected loss EL$=\PD\cdot \LGD$ for some given LGD estimate and average default probability $\PD$ for that asset class. The asset correlation can then be inferred from the standard deviation of the loss rates as follows. At first, a specific distribution, e.g.\ a beta distribution, is fitted to the first two moments of the loss rate data, for which we then estimate the quantile at a given level, say $q=99.9\%$. This quantile reflects both the expected and unexpected loss and by subtracting the expected loss, we obtain the stand-alone UL. Finally, the correlation coefficient $\rho$ is determined such that the UL under the Basel regulatory framework agrees with the empirical UL, obtained from the beta distribution.

\section{Supplementary Material}\label{app supplementary material}

\begin{landscape}

\begin{table}[H]
    \centering
    
%\begin{adjustbox}{angle=90}

\scalebox{0.7}{
    \begin{tabular}{lrrrrrrrrrrrrrrrrrr}
&AAA&AA+&AA&AA-&A+&A&A-&BBB+&BBB&BBB-&BB+&BB&BB-&B+&B&B-&Cs&D\\
\hline
AAA &96.79 & 2.71& 0.42& 0.00& 0.00& 0.00& 0.01& 0.00& 0.00& 0.00& 0.00& 0.07& 0.00& 0.00& 0.00& 0.00& 0.00& 0.00\\
AA+& 6.45& 85.16& 6.61& 1.77& 0.00& 0.00& 0.00& 0.00& 0.00& 0.00& 0.00& 0.00& 0.00& 0.00& 0.00& 0.00& 0.00& 0.00\\
AA &0.00& 6.22& 85.17& 6.74& 0.52& 0.42& 0.10& 0.52& 0.00& 0.00& 0.00& 0.31& 0.00& 0.00& 0.00& 0.00& 0.00& 0.00\\
AA- &0.00& 0.00& 7.82& 83.45& 7.16& 0.17& 0.50& 0.17& 0.00& 0.17& 0.44& 0.00& 0.00& 0.11& 0.00& 0.00& 0.00& 0.00\\
A+ &0.00& 0.00& 0.07& 13.35& 73.28& 9.30& 2.03& 1.12& 0.14& 0.63& 0.07& 0.00& 0.00& 0.00& 0.00& 0.00& 0.00& 0.01\\
A &0.00& 0.00& 0.00& 1.15& 12.33& 77.29& 5.71& 1.68& 0.77& 0.96& 0.10& 0.00& 0.00& 0.00& 0.00& 0.00& 0.00& 0.01\\
A- &0.00& 0.00& 0.00& 0.00& 0.94& 11.47& 77.82& 6.94& 0.41& 1.57& 0.67& 0.16& 0.00& 0.00& 0.00& 0.00& 0.00& 0.02\\
BBB+& 0.00& 0.00& 0.00& 0.00& 0.00& 2.16& 12.39& 70.86& 11.24& 2.41& 0.60& 0.24& 0.06& 0.00& 0.00& 0.00& 0.00& 0.04\\
BBB &0.00& 0.00& 0.00& 0.00& 0.00& 0.00& 1.87& 16.60& 68.05& 11.16& 0.99& 0.11& 0.00& 0.50& 0.22& 0.11& 0.33& 0.06\\
BBB-& 0.00& 0.00& 0.00& 0.00& 0.00& 0.00& 0.00& 0.93& 14.94& 74.69& 6.50& 2.13& 0.27& 0.08& 0.15& 0.12& 0.08& 0.11\\
BB+& 0.00& 0.00& 0.00& 0.00& 0.00& 0.00& 0.00& 0.00& 0.54& 20.57& 66.38& 9.93& 1.14& 0.18& 0.06& 0.00& 1.02& 0.18\\
BB &0.00& 0.00& 0.00& 0.00& 0.00& 0.00& 0.00& 0.00& 0.00& 0.78& 14.20& 70.80& 11.15& 1.80& 0.68& 0.15& 0.05& 0.40\\
BB- &0.00& 0.00& 0.00& 0.00& 0.00& 0.00& 0.00& 0.00& 0.00& 0.00& 1.03& 10.55& 73.79& 11.47& 1.19& 0.46& 0.61& 0.90\\
B+ &0.00& 0.00& 0.00& 0.00& 0.00& 0.00& 0.00& 0.00& 0.00& 0.03& 0.03& 0.92& 10.20& 68.70& 15.30& 2.50& 0.85& 1.46\\
B& 0.00& 0.00& 0.00& 0.00& 0.00& 0.00& 0.00& 0.00& 0.00& 0.00& 0.00& 0.00& 0.61& 13.50& 70.77& 9.84& 2.91& 2.38\\
B- &0.00& 0.00& 0.00& 0.00& 0.00& 0.00& 0.00& 0.00& 0.00& 0.00& 0.00& 0.00& 0.00& 2.43& 15.07& 66.85& 8.05& 7.59\\
Cs &0.00& 0.00& 0.00& 0.00& 0.00& 0.00& 0.00& 0.00& 0.00& 0.00& 0.00& 0.00& 0.00& 0.74& 1.81& 13.90& 32.08& 51.47\\
D &0.00& 0.00& 0.00& 0.00& 0.00& 0.00& 0.00& 0.00& 0.00& 0.00& 0.00& 0.00& 0.00& 0.00& 0.00& 0.00& 0.00& 100.00\\
\hline
    \end{tabular}
}
%\end{adjustbox}
\caption{Normalized average one-year sovereign foreign currency transition matrix (transition probabilities in \%).}\label{tab transition matrix}
\end{table}
\end{landscape}

\begin{figure}[htb]
\begin{center}
    \subfigure[Rating distribution for AFDB, CDB and TDB.]{
         \includegraphics[width=0.6\textwidth]{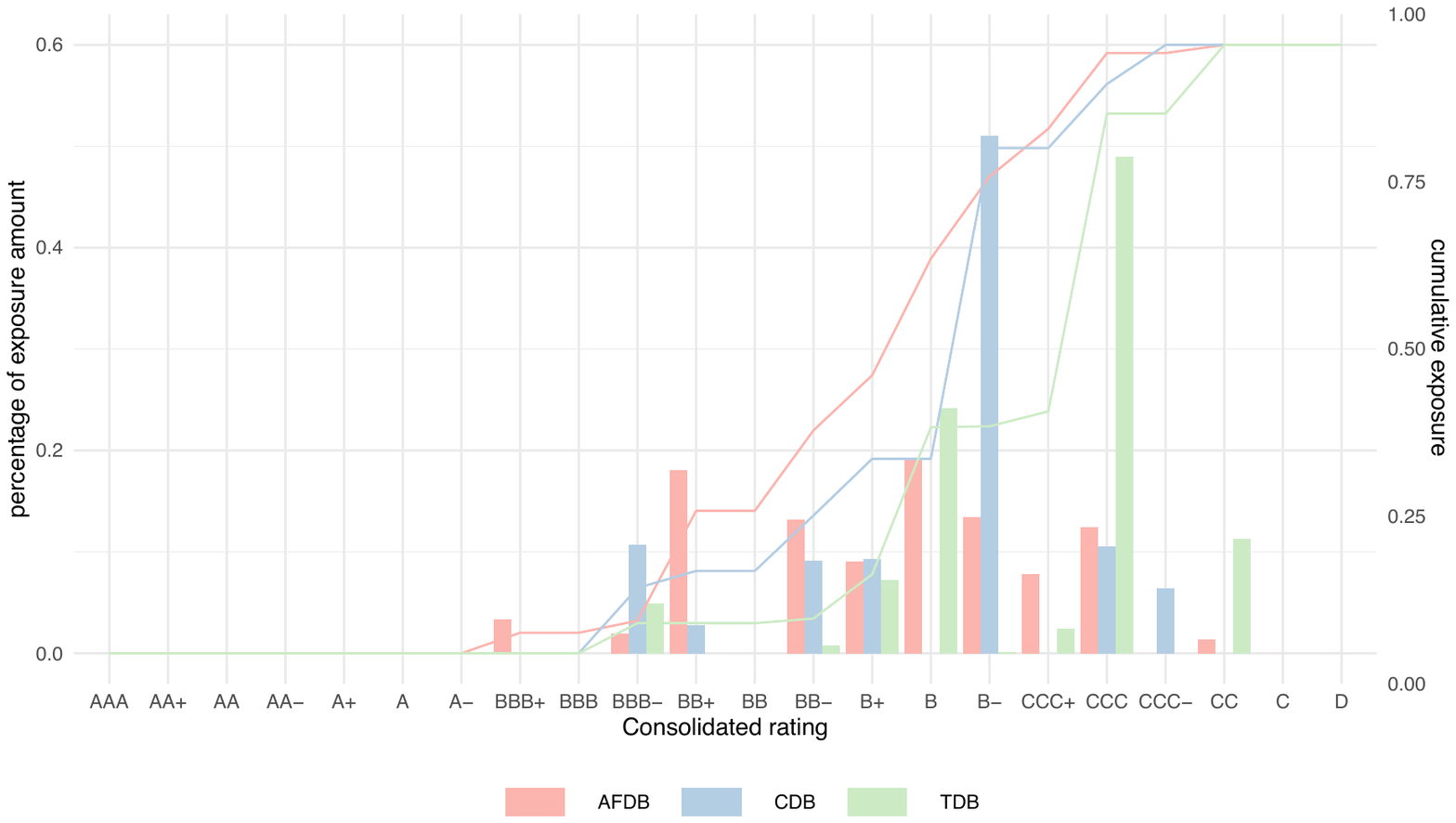} \label{fig PD AFDB CDB and TDB}}
    \subfigure[Rating distribution for BOAD, CABEI and EADB.]{
         \includegraphics[width=0.6\textwidth]{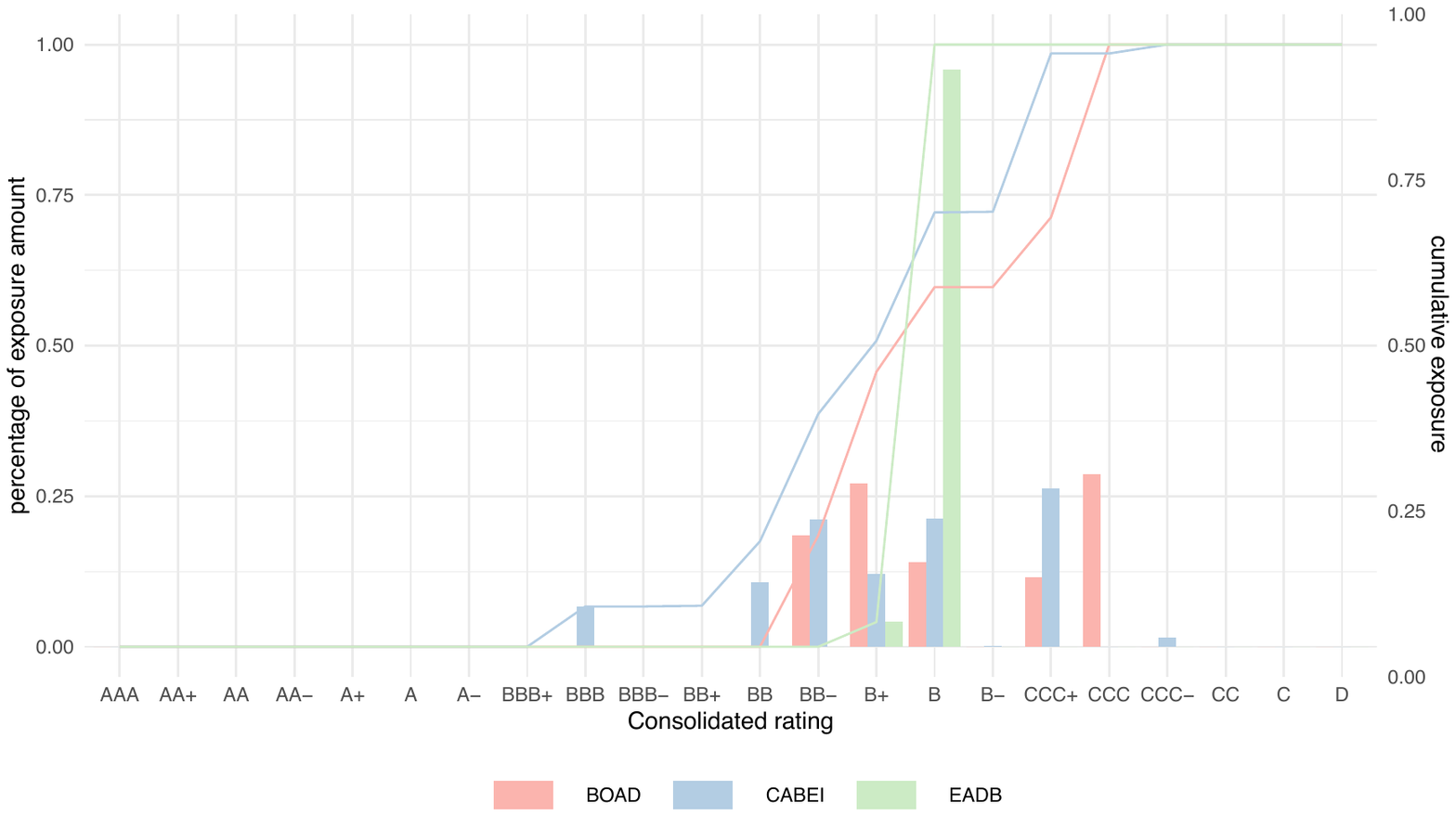} \label{fig PD BOAD CABEI and EADB}}
    \subfigure[Rating distribution for CAF, EBRD, and IDB.]{
         \includegraphics[width=0.6\textwidth]{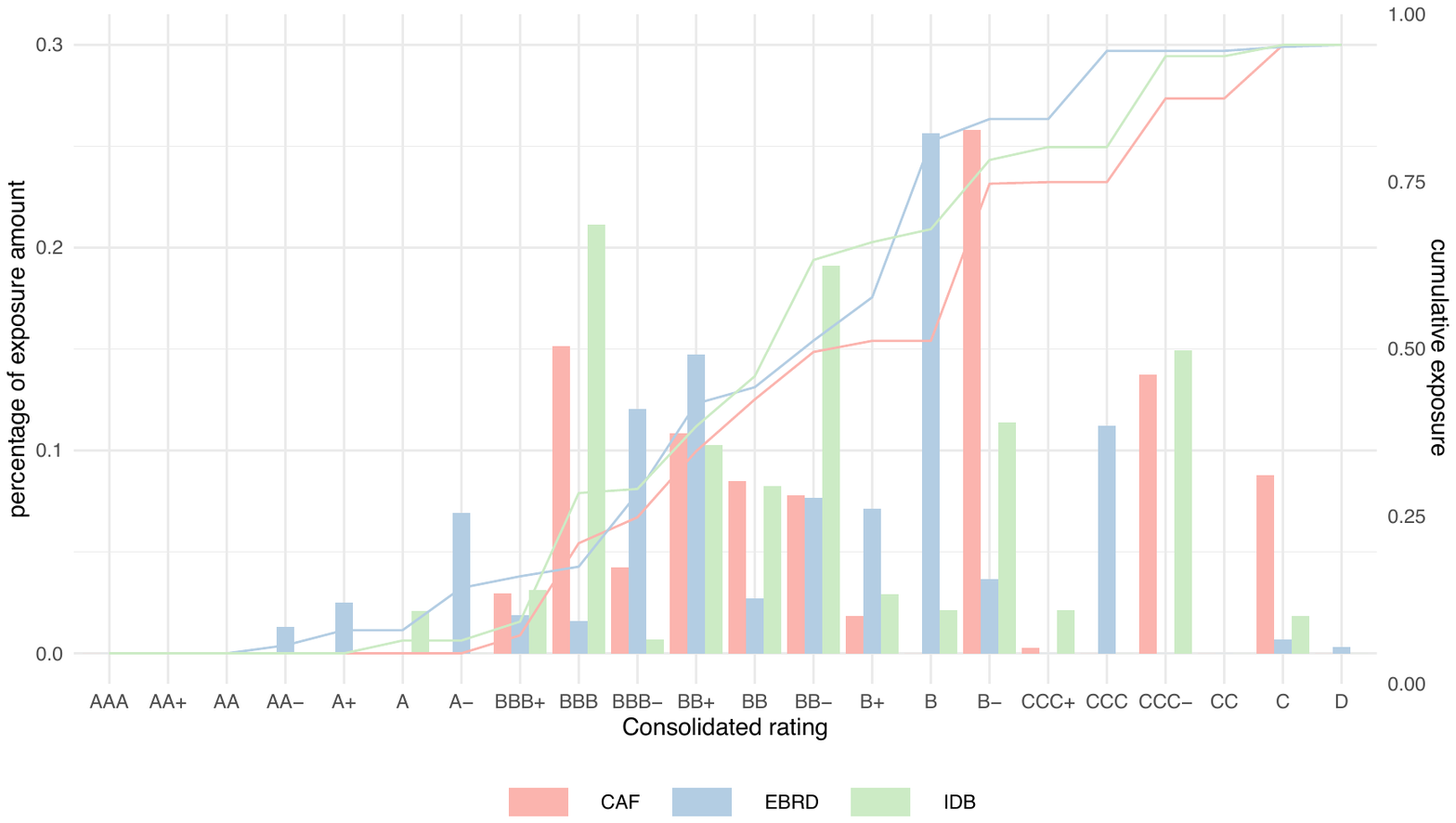} \label{fig PD CAF IDB and EBRD}}
\caption{Rating distribution in MDB portfolios. The figure shows the rating distribution of the sovereign loan portfolios of various MDBs as of 2022.} \label{fig rating distributions other MDBs}
\end{center}
\end{figure}

\begin{figure}[htb]
\begin{center}
        \subfigure[Maturity distribution for BOAD, EADB and IBRD.]{
         \includegraphics[width=0.4\textwidth]{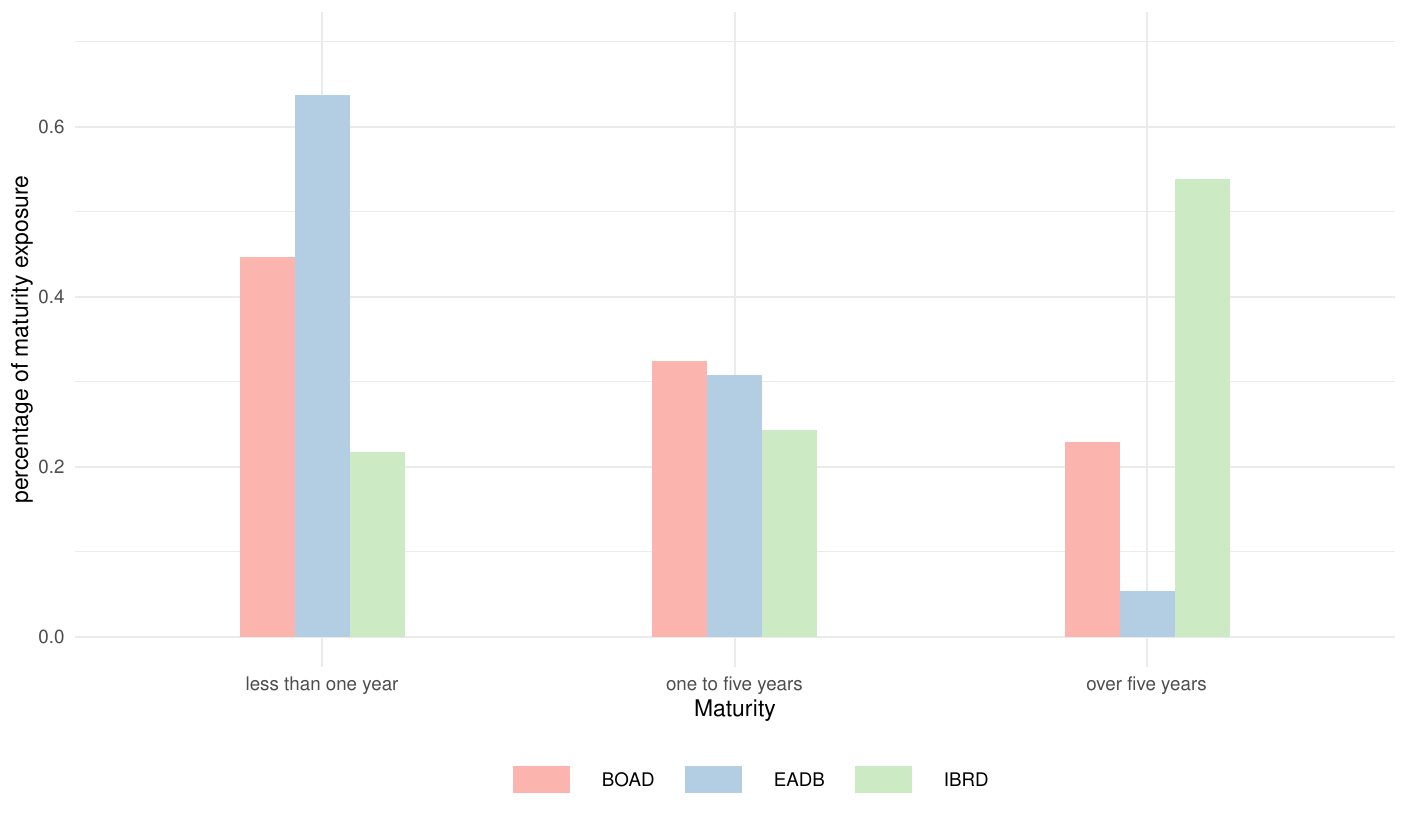} \label{fig maturity BOAD EADB and IBRD}}
    \subfigure[Maturity distribution for ADB and IDB.]{
         \includegraphics[width=0.4\textwidth]{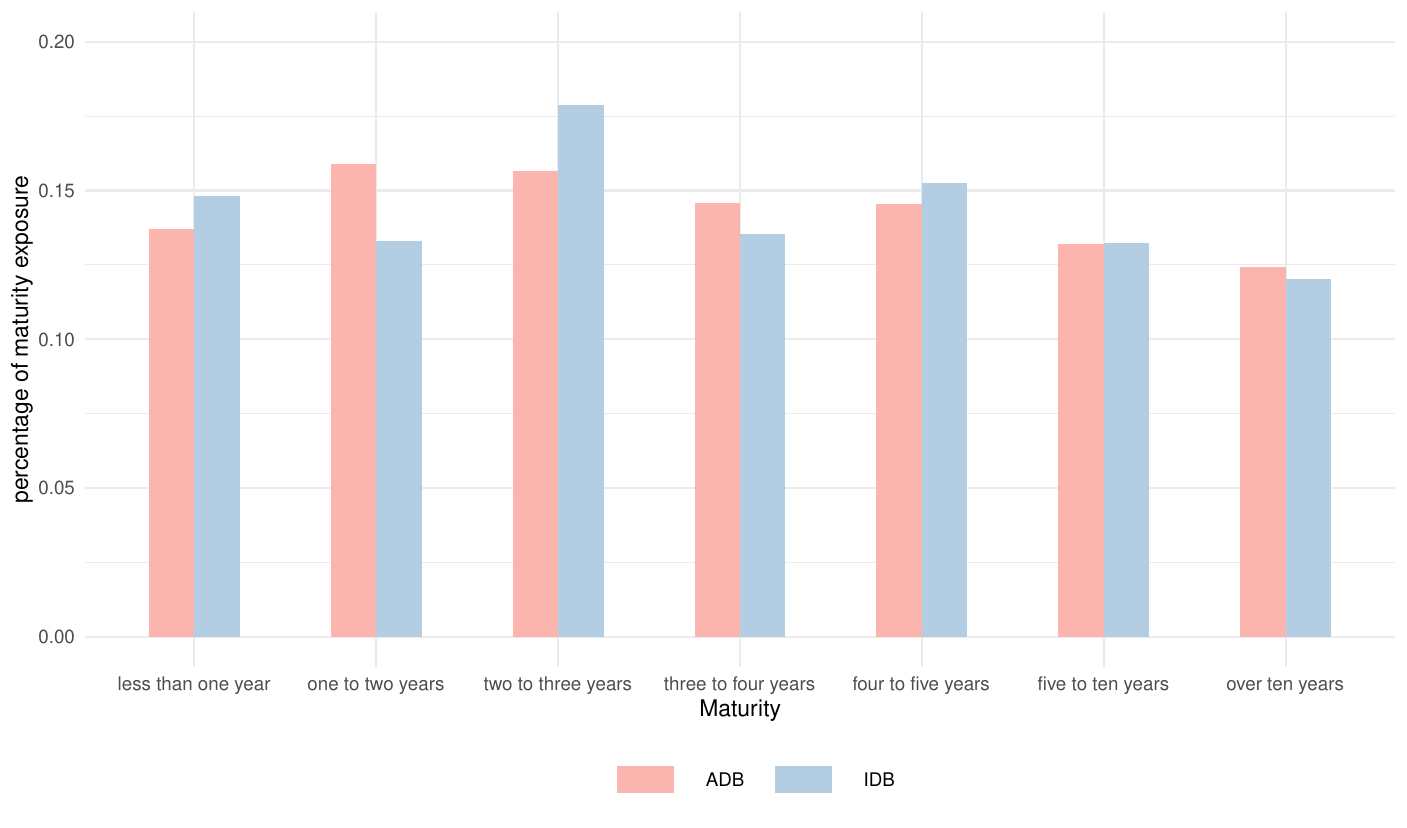} \label{fig maturity ADB and IDB}}
\caption{Maturity distribution in MDB portfolios. The figure shows the maturity distribution of the sovereign loan portfolios of various MDBs as of 2022. Note that maturity information on loan portfolios is not available for all MDBs.} \label{fig maturity distributions other MDBs}
\end{center}
\end{figure}

\begin{figure}[htb]
\begin{center}
       \subfigure[Exposure distribution for AFDB, EBRD, IDB and TDB.]{
         \includegraphics[width=0.4\textwidth]{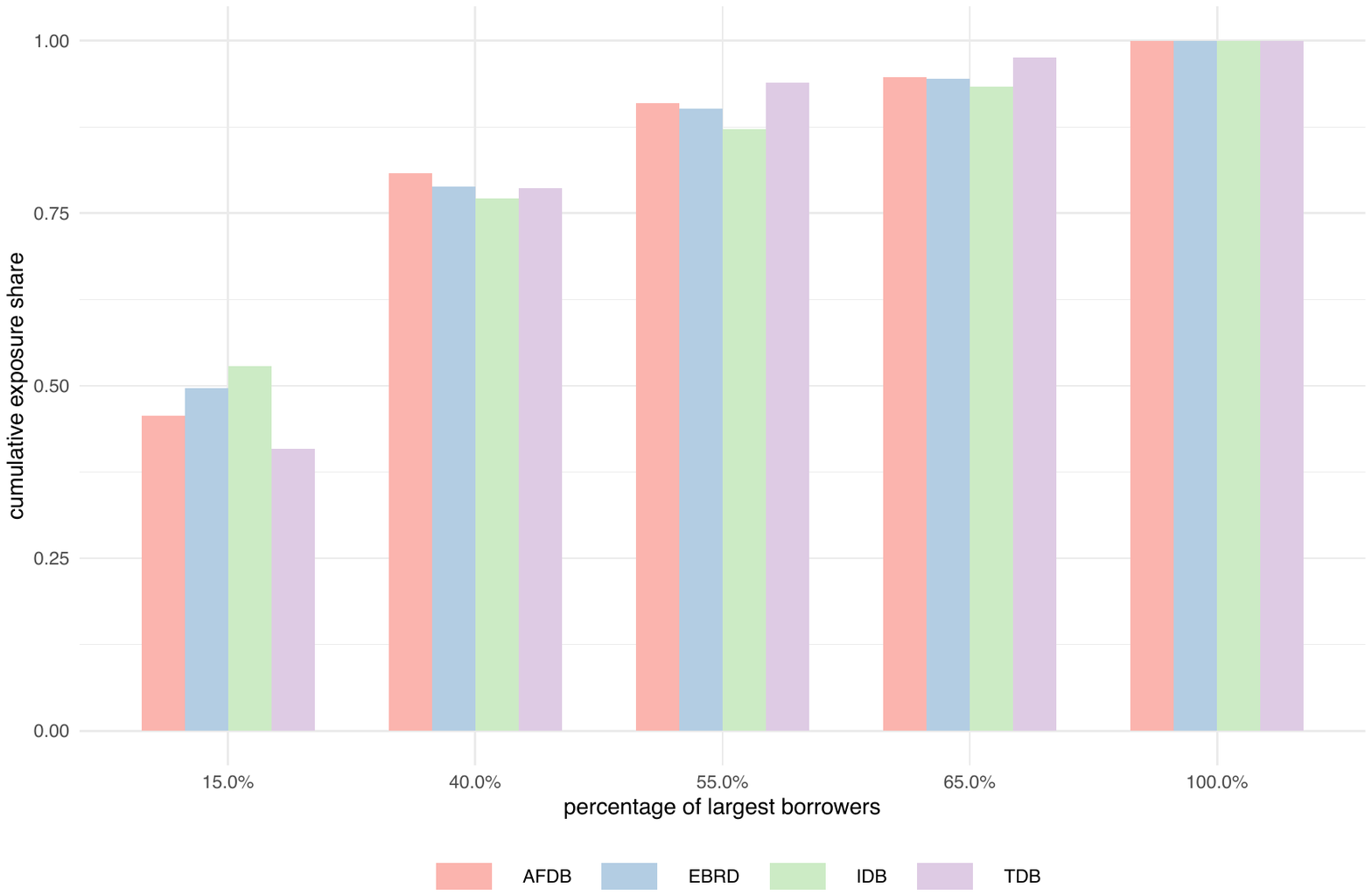} \label{fig exposure distribution AFDB EBRD IDB TDB}}
        \subfigure[Exposure distribution for CAF and CDB.]{
         \includegraphics[width=0.4\textwidth]{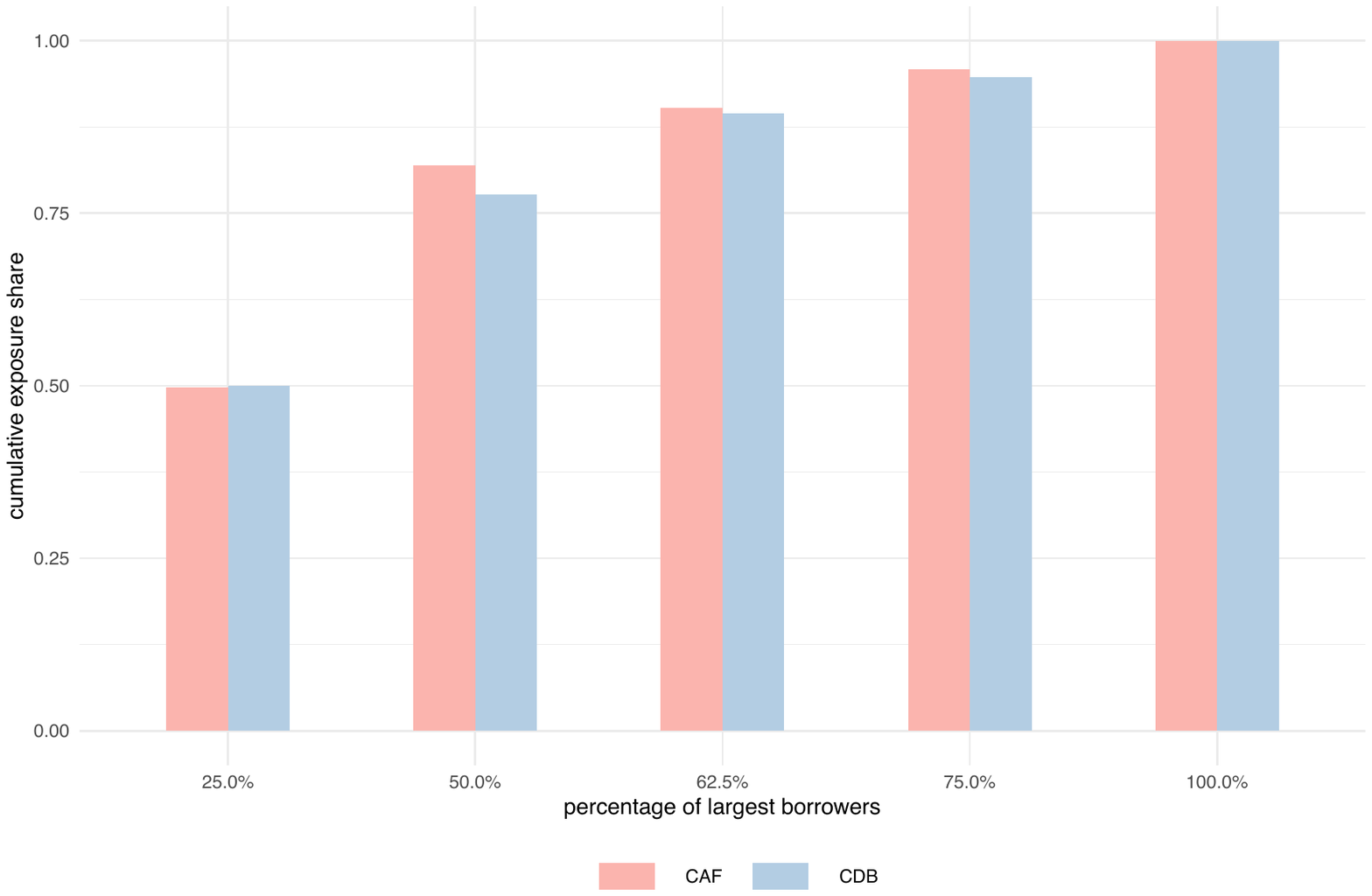}\label{fig exposure distribution CAF CDB}}
    \subfigure[Exposure distribution for BOAD, CABEI and EADB.]{
         \includegraphics[width=0.4\textwidth]{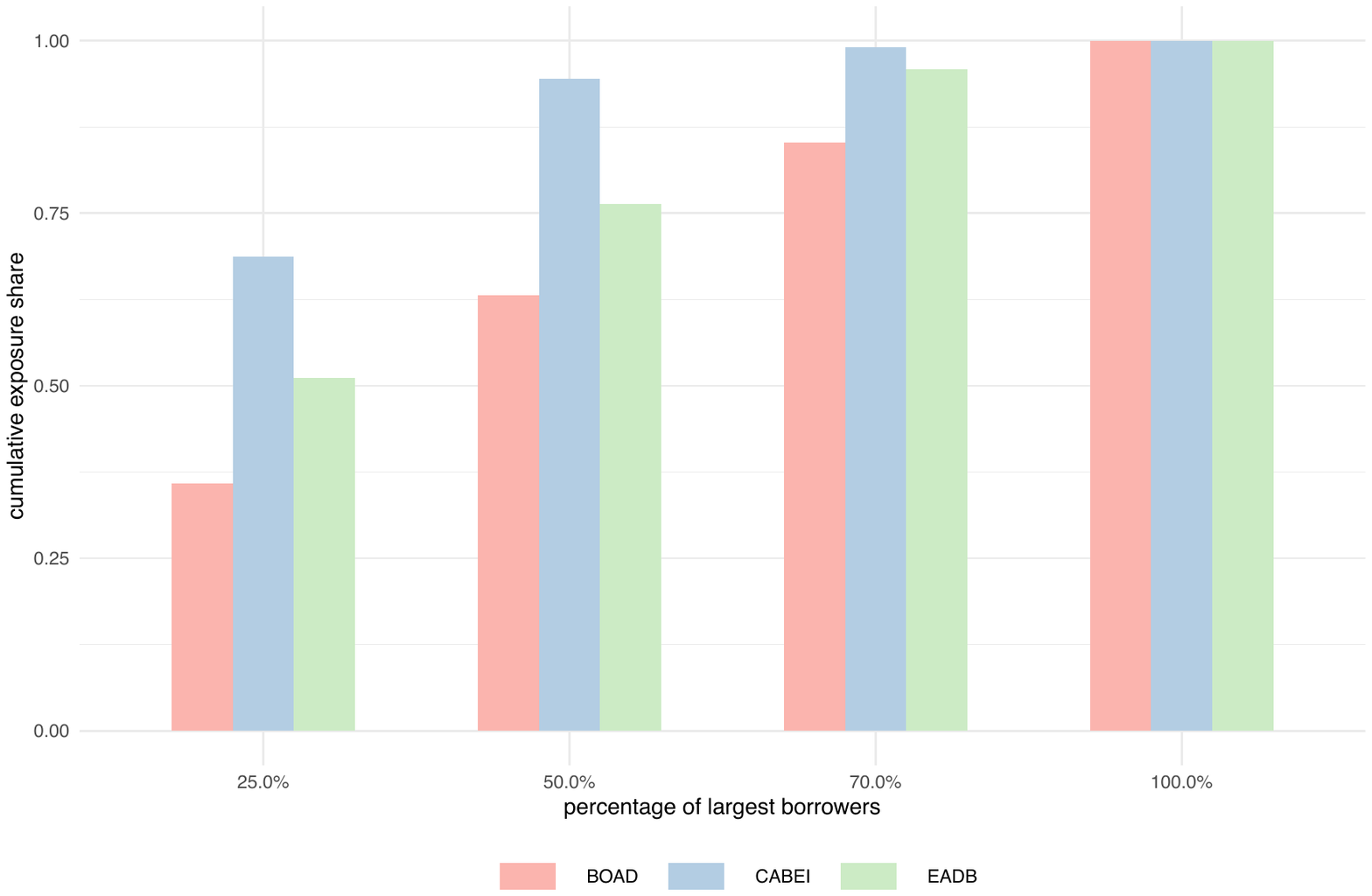} \label{fig exposure distribution CABEI BOAD and EADB}}
\caption{Exposure distribution in MDB portfolios. The figure shows the exposure distribution of the sovereign loan portfolios of various MDBs as of 2022. } \label{fig exposure distributions other MDBs}
\end{center}
\end{figure}

\end{document}